\documentclass[12pt,a4paper]{article}
\usepackage{amssymb, amsmath, amsthm}
\usepackage{graphicx}
\usepackage{cite}

\usepackage{hyperref}

\def\be{\begin{equation}}
	\def\ee{\end{equation}}
\def\bea{\begin{eqnarray}}
	\def\eea{\end{eqnarray}}

\newcommand{\ba}{\begin{eqnarray}}
	\newcommand{\ea}{\end{eqnarray}}
\newcommand{\ban}{\begin{eqnarray*}}
	\newcommand{\ean}{\end{eqnarray*}}

\theoremstyle{definition}

		\title{\bf Properties of a deformed Myers-Perry black hole}
\author{ Shohreh Abdolrahimi $^a$\footnote{sabdolrahimi@cpp.edu}, Matin Tavayef$^b$\footnote{mtavayef@mun.ca} \\ \\
	\small \sl $^a$Department of Physics and Astronomy, California State Polytechnic University, \\ \small \sl 
	Pomona, CA 91768, USA  \\
	\small \sl $^b$Faculty of Science, Theoretical Physics, Memorial University of Newfoundland \\ \small \sl  St. John's, NL  A1C 5S7, Canada } 

\date{}
\begin{document}
	\maketitle
	\begin{abstract}
	Considering an exact solution of the five-dimensional Einstein equations in vacuum space, which represents a distorted Myers-Perry black hole with a single angular momentum, we investigate how the distortion affects the horizon surface of this black hole. We illustrate a special case where we have a bumpy deformed black hole horizon that feels the presence of the external sources. However, the ergosphere is oblivious to the presence of external sources.
	\end{abstract}
	
\section{Introduction}
The study of the exact solutions of Einstein's equations describing black holes and their features is at the core of general relativity. From a theoretical perspective, they are remarkable, considering that they shed light on the characteristics of gravitational theory in extreme settings and, as well as being astronomical objects, are engaged in, or act as a background, for a variety of physical processes.
Nonetheless, the majority of the exact solutions of Einstein's equations describe a single, isolated black hole.
For instance, the Kerr-Newmann family of solutions represents the unique stationary charged black hole in the asymptotically flat four-dimensional spacetime.
 However, from the theoretical viewpoint, the study of the isolated black hole would be equivalent to studying the properties of an isolated charged object. As it is evident from this analogy, investigating the characteristics of distorted black holes is exactly as important as studying the interaction of charged objects. Furthermore, the majority of astrophysical scenarios indicate that the black hole is not isolated but rather interacts with an outside matter distribution. These systems are generally dynamical and due to their complexity, we often need to employ numerical or perturbative approaches. However, exact solutions exist in idealized scenarios where a specific form of the external matter is assumed.
Geroch and Hartle developed \cite{DistortedBH} a method with exact solutions for studying the non-isolated black holes interacting with external matter sources. 
They studied the characteristics, thermodynamical behaviour and Hawking radiation of general static black holes in four-dimensional spacetime in the presence of external matter fields.
They studied solutions to the vacuum static Einstein equations that include a regular horizon and lack singularities in the outer communication domain, while being asymptotically non-flat when viewed globally. This solution called a distorted black hole, represents a local solution, valid only in the neighborhood of the black hole
horizon. These solutions also contain information related to external matter fields, although they don't include the external matter, since they represent solutions of the vacuum Einstein equations. 
A significant benefit of these solutions is that they apply to wide classes of external matter, with restrictions only arising from certain regularity conditions. A distorted black hole can be viewed as a potential approximation for dynamical black holes that settle on a timescale much shorter than that of the external matter, or for equilibrium systems where black holes and matter move in a quasi-stationary state.
 Several local static vacuum distorted black hole solutions were analyzed explicitly
\cite{Israel64,dis4,doroshkevich1965,Erez59,Mysak}.
 Furthermore, \cite{Tomimatsu},\cite{Kerbh230} extended this to the rotating case by using solitonic techniques to describe Kerr black holes in arbitrary axisymmetric external gravitational fields. In addition, within the frameworks of classical general relativity and dilaton gravity, four-dimensional distorted charged black holes have been studied.
In \cite{Fairhurst} and \cite{yazadjiev2001}, static solutions have been constructed with dilaton fields. Additionally, a Kerr-Newman black hole surrounded by an external gravitational field was obtained in\cite{Breton1998}. In contrast to their non-distorted asymptotically flat counterparts, it was found that four dimensional distorted black holes can have a regular horizon with toroidal topology, with the exception of the spherical one\cite{Peters}, \cite{Xanthopoulos}. Also, higher dimensional distorted static vacuum black hole solutions
with a spherical horizon, such as distorted 5D
Schwarzschild-Tangherlini black hole, and a static electrically
charged solution with spherical horizon (distorted 5D
Reissner-Nordstr\"{o}m black hole) have been studied before\cite{DBHV,ElCh2014}.
Furthermore, other five dimensional distorted black holes such as static black ring and a distorted static black hole with a bubble has been studied in \cite{Blackring,Tavayef2022}.

Distorted black holes are more generic solutions of the Einstein equations. This feature makes them an intriguing topic to study.
Compared them to isolated black holes, we can get deeper insights into the properties of black holes. Several studies were conducted to examine the impact of external matter fields on the properties of isolated black holes, specifically how they are deformed. We may ask which features of the black holes stay unchanged regardless of the deformations. Certain characteristics, by contrast, are a consequence of the idealized symmetry of undistorted black holes. A number of studies have explored the impact of external matter fields on the properties of isolated black holes\cite{papadopoulos1984,Breton1998, hennig2010, Ansorg, Ashtekar1, Ashtekar3, ElCh2014}. For example, it is worth mentioning that the ratios between the mass and the angular momentum of the distorted black holes can be different 
from their corresponding asymptotically flat cases which lack external matter fields. 
For instance, for Kerr black hole, the existence of an event horizon is dependent on the mass $M$ and the angular momentum $J$ and they should satisfy the inequality  $J/M^2 \leq 1$ (this inequality holds even for dynamical, axisymmetric black holes)
\cite{Dain:2006wb}.
Interestingly, when a Kerr black hole is surrounded by an external matter field, this ratio can surpass one and become infinitely large \cite{ansorg2005},\cite{Ansorg2}.  The geometry of the horizon of a distorted black hole can differ significantly from that of isolated black holes. Even in the four-dimensional static scenario, their horizon surfaces are axisymmetric and can exhibit significant elongation or flattening.
The horizon of the distorted Reissner-Nordström black hole is only shown to be spherically symmetric in the extremal limit. \cite{Booth2}.

The purpose of this work is to analyze the properties of a distorted five-dimensional Myers-Perry black hole surrounded by an external gravitational field, which was generated in \cite{MPBH}. According to Geroch and Hartle's interpretation, the solution describes a Myers-Perry black hole locally when an external matter field is present. Actually, this solution has been derived by applying a two-fold B\"{a}cklund transformation, a method proposed by Neugebauer for solving four-dimensional stationary and axisymmetric problems.
We considered a five-dimensional vacuum Weyl solution as a seed which describes a regular region in spacetime with a static external matter distribution.
The result after application of the two-fold B\"{a}cklund transformation is a vacuum solution describing a local deformed black hole rotating around one of its symmetry axes. 

The paper is organized as follows: We review the Myers-Perry solution in Section 2, and the distorted Myers-Perry black hole in Section 2.1. In Section 3, we analyze the deformations of the horizon surface and the curvature of the horizon sections. 
\section{Myers-Perry Solution, Review}
The Myers-Perry solution \cite{Myers} is a stationary solution to the Einstein equations. A solution which describes a rotating black hole with spherical horizon in a D-dimensional vacuum spacetime.
The Kerr black hole in $D=4$ is a particular case of the Myers-Perry family.
Additionally, the solution is axisymmetric and contains $N$ spacelike Killing fields, where $N$ is the integer component of $(D-1)/2$. The number of rotational axes is denoted by $N$ in this instance. Generally, the solution is described by $N$ independent angular momenta. In this paper, we focus on a locally distorted Myers-Perry black hole in $D=5$ with a single angular parameter.

The Myers-Perry black hole solution in 5-dimensions is in general characterized by two rotation parameters $a_1$ and $a_2$. These parameters are associated with the angular momentum directed along the axis of rotation. When one of the angular momenta parameter, for example, $a_1$ vanishes, it is more convenient to use the prolate spheroidal coordinates $x$ and $y$. The Myers-Perry black hole with a single non-vanishing rotation parameter acquires the form \cite{PRD2006}
\begin{eqnarray}\label{Myers-PerryP}
	ds^2&=& -\frac{x-1-\alpha^2(1-y)}{x+1+\alpha^2(1+y)}
	\left(dt + 2\sigma^{1/2}\alpha
	\frac{(1+\alpha^2)(1-y)}{x-1-\alpha^2(1-y)} d\psi\right)^2
	\nonumber \\[2mm]
	&+& \sigma\frac{(x-1)(1-y)(x+1+\alpha^2(1+y))}{x-1-\alpha^2(1-y)} d\psi^2
	+ \sigma(x+1)(1+y)d\phi^2
	\nonumber \\[2mm]
	&+& \frac{\sigma}{2}(x+1+\alpha^2(1+y))
	\left( \frac{dx^2}{x^2-1}+\frac{dy^2}{1-y^2}\right).
\end{eqnarray}
In this solution, $x$ and $y$ are the prolate spheroidal coordinates, taking the range $x\geq 1$, and $-1\leq y \leq 1$. The physical infinity corresponds to the limit $x\rightarrow \infty$ and the black hole horizon is located at $x=1$. The axes for the spacelike Killing fields $\frac{\partial}{\partial\phi}$ and $\frac{\partial}{\partial\psi}$ are located at $y=-1$ and $y=1$ respectively. The solution is characterized with two parameters $\sigma$ and $\alpha$. The parameter $\alpha$ is related to $(m, a_2)$ in Boyer-Lindquist coordinates by $\alpha^2 = a^2_2/(m-a^2_2)$. The ADM mass $M$, the angular momentum $J$ and the angular velocity $\Omega$ of the solution are given in terms of the parameters $\{\sigma, \alpha\}$ by
\begin{eqnarray}\label{M_MP}
	M=\frac{3\,\pi}{2}\,\sigma (1+\alpha^2), \quad~~~J = 2\pi\sigma^{\frac{3}{2}}\alpha (1+\alpha^2),\quad~~~  \Omega = \frac{\alpha}{2\sqrt{\sigma}(1+\alpha^2)}
\end{eqnarray}
Obviously, when the rotation parameter $\alpha$ vanishes, the solution has to reduce to the five-dimensional Schwarzschild-Tangherlini black hole presented in terms of the prolate spheroidal coordinates \cite{abdolrahimi2014B}. 
\subsection{Distorted 5D Myers-Perry black hole}

The distorted 5D Myers-Perry black hole with one rotation parameter was constructed in \cite{abdolrahimi2014B} using the 2-fold B\"{a}cklund transformation on the distorted Minskowski spacetime as a background. The solution can be presented in the following form
\begin{eqnarray}
	ds^2&=& -\frac{x-1-\hat{a}^2(1-y)}{x+1+\hat{a}^2(1+y)}e^{2(\widehat{U} + \widehat{W})}
	\left(dt -\omega d\psi\right)^2
	+\frac{x+1+\hat{a}^2(1+y)}{x-1-\hat{a}^2(1-y)}e^{-2W} d\psi^2
	\nonumber \\[2mm]
	&+& e^{-2U}d\phi^2 + C_1\left[x+1+\hat{a}^2(1+y)\right]e^{2(\hat{\gamma} - \widehat{W})}
	\left( \frac{dx^2}{x^2-1}+\frac{dy^2}{1-y^2}\right), \label{metric_dist} \\
\end{eqnarray}
where
\ba
\gamma &=& \gamma_0 + \hat{\gamma}, \nonumber \\
\gamma_0 &=& \frac{1}{2}\ln(x^2-1) - \frac{1}{2}\ln(x^2-y^2) +  \frac{1}{2}\ln(y+1), \nonumber \\
\hat{\gamma} &=& \sum^{\infty}_{n,k=1}\frac{nk}{n+k}\left(a_na_k + a_nb_k + b_nb_k\right)R^{n+k}\left(P_n P_k - P_{n-1}P_{k-1}\right) \nonumber \\
&+& \frac{1}{2}\sum^{\infty}_{n=1}(2a_n+b_n)\sum^{n-1}_{k=0}(-1)^{n-k+1}(x+y)R^k P_k\left(\frac{xy}{R}\right) \nonumber \\
&-& \frac{1}{2}\sum^{\infty}_{n=1}(a_n+2b_n)\sum^{n-1}_{k=0}(x-y)R^k P_k\left(\frac{xy}{R}\right) \nonumber \\
&-&\frac{1}{2}\sum^{\infty}_{n=0}(a_n - b_n)R^n P_n\left(\frac{xy}{R}\right),\\
\omega &=& -2\sqrt{\sigma}\frac{(x-y)\hat{a}e^{-2\widehat{W}-\widehat{U}}}{(x-1)-(1-y)\hat{a}^2} + C_\omega, \label{omega_dist} \\[2mm]
\hat{a} &=& \alpha \exp{\left[\sum^{\infty}_{n=1}(a_n+2b_n)\sum^{n-1}_{k=0}(x-y)R^k P_k\left(\frac{xy}{R}\right)\right]},
\ea
and 
\begin{eqnarray}\label{UW_dist}
	\widehat{U} &=& \sum^{\infty}_{n=0}a_nR^n P_n\left(\frac{xy}{R}\right), \quad ~~~
	\widehat{W} = \sum^{\infty}_{n=0}b_nR^n P_n\left(\frac{xy}{R}\right)\,. 
\end{eqnarray} 
The solution has two integration constants $C_1$ and $C_\omega$
To prevent pathological behavior, the two integration constants in the solution, $C_1$ and $C_\omega$, should be chosen appropriately. Thus, we have
\begin{eqnarray}\label{C_omega}
	&&C_\omega = 2\sqrt{\sigma}\alpha\exp{\left[-\sum^\infty_{n=0}(a_n + 2b_n)\right]}\, \\
	&&C_1 = \frac{\sigma}{2}. 
\end{eqnarray} 

Avoiding the conical singularity leads to the following condition on the multiple moments
\begin{eqnarray}\label{conicalmp}
	\sum^\infty_{n=0}(b_{2n} - a_{2n}) + 3\sum^\infty_{n=0}(a_{2n+1} + b_{2n+1})= 0~.
\end{eqnarray}
The solution has two limits. When the rotation parameter vanishes, the solution reduces to the 5D distorted Schwarschild-Tangherlini black hole, which is obtained in \cite{DBHV}. The other limit is obtained when all the multiple moments vanish, i.e., $a_n = 0$, $b_n = 0$ for every $n$, and the solution reduces to the asymptotically flat Myers-Perry black hole with a single rotation parameter $(\ref{Myers-PerryP})$. 
The prolate spheroidal coordinates $(x,y)$ are related to $(\eta,\theta)$ by $x = \eta$, $y = \cos\theta$. We also have $\sigma = r_0^2/4$.  The horizon is encompassed by an ergoregion, where the Killing field $\partial/\partial t$ is spacelike. The Komar mass $M_H$ and angular momentum $J_H$ associated with the black hole horizon are given by
\begin{eqnarray}\label{Komar}
	M_{H} &=&  - {3\over 32\pi} \int_{H} \star d\xi =\frac{3\,\pi}{2}\,\sigma (1+\alpha^2),  \label{M_dist}, \nonumber \\
	J_H &=&  {1\over 16\pi} \int_{H} \star d \zeta=2\pi\sigma^{\frac{3}{2}}\alpha (1+\alpha^2)\exp{\left[-\sum^\infty_{n=0}(a_n + 2b_n)\right]} \label{J_dist},
\end{eqnarray}
The mass has the same form as the asymptotically flat undistorted Myers-Perry black hole, while the angular momentum differs by the exponential factor. Since the solution is not asymptotically flat, the $ADM$ mass is not defined. We have the following relation $J^2_H/M^3_H$ for the distorted Myers-Perry black hole 
\begin{eqnarray}
	\frac{27\pi}{32}\frac{J^2_H}{M^3_H} = \frac{\alpha^2}{1+\alpha^2}\exp{\left[-2\sum^\infty_{n=0}(a_{n} + 2b_{n})\right]}.
\end{eqnarray}

\section{Horizon Surface}
In this paper, we mainly study the effect of the distortion fields on the black hole horizon surface, which is defined by $t=const$ and $x=1$. The horizon surface is the constant time slice of the horizon. The metric on the horizon surface is given by the expressions \cite{MPBH}
\begin{eqnarray}\label{metric_H}
	ds^2_{H} &=& \frac{4\sigma(1+\alpha^2)^2}{2 + {\hat a}^2(1+y)}(1-y)e^{-2{\widehat W}}d\psi^2 + 2\sigma(1+y)e^{-2{\widehat U}}d\phi^2   \nonumber \\[2mm]
	&+& \frac{\sigma}{2}\left[2 + {\hat a}^2(1+y)\right]e^{2({\hat\gamma}-{\widehat W})}\frac{dy^2}{1-y^2}. 
\end{eqnarray}
Here, the functions ${\hat a}$, ${\widehat W}$, ${\widehat U}$ and ${\hat\gamma}$ are given by
\begin{eqnarray}
	&&{\widehat U} = {\sum^\infty_{n=0}a_n\,y^n}, \quad~~~{\widehat W} = {\sum^\infty_{n=0}b_n\,y^n},  \nonumber \\[2mm]
	&&{\hat a} = \alpha\exp{\left[-\sum^\infty_{n=0}(a_n + 2b_n)(y^n-1)\right]},  \nonumber \\[2mm]
	&&{\hat\gamma} = {\sum^\infty_{n=0}(a_n + 2b_n)y^n -\frac{3}{2}\sum^\infty_{n=0}(a_{2n} + b_{2n}) - \frac{1}{2}\sum^\infty_{n=0}(b_{2n+1} - a_{2n+1})} \,\nonumber.
\end{eqnarray}
Using the angular coordinate $y=\cos \theta$, $0\leq\theta\leq {\pi}$, the metric of the horizon surface takes the form:
\begin{eqnarray}\label{HorizonM}
	ds^2_{H} &=& \frac{4\sigma(1+\alpha^2)^2}{1 + {\hat a}^2(\theta)\cos^2(\frac{\theta}{2})}\sin^2(\frac{\theta}{2}) e^{-2{\widehat W}(\theta)}d\psi^2 + 4\sigma \cos^2(\frac{\theta}{2}) e^{-2{\widehat U}({\theta})}d\phi^2 \nonumber \\[2mm]
	&+& \sigma\left[1 + {\hat a}^2({\theta})\cos^2(\frac{\theta}{2})\right]e^{2({\hat\gamma}(\theta)-{\widehat W}(\theta))} d\theta^2.
\end{eqnarray}
\noindent
We see that in the limit where the distortion multiple moments $a_n$'s and $b_n$'s, and the rotation parameter $\alpha$ vanish, we obtain the metric on the 3-sphere in Hopf coordinates with radius $R = 2\sqrt{\sigma}$, 
\begin{eqnarray}
	ds^2_{H} = 4\sigma\left[d\lambda^2 + \sin^2\lambda d\psi^2 + \cos^2\lambda d\phi^2\right],\label{Horizonnorotation}
\end{eqnarray}
where $\lambda=\theta/2$, $\{0\leq\lambda\leq \frac{\pi}{2}, 0<\phi<2\pi, 0<\psi<2\pi\}$.   
In the absence of distortions the metric of the horizon surface is 
\be
ds_{H}^{2}=4\sigma[f(\lambda)d\lambda^{2}+\frac{(1+\alpha^2)^2\sin^{2}\lambda}{f(\lambda)}d\psi^{2}+\cos^{2}\lambda d\phi^{2}\biggl],\label{HorizonUndis}
\ee
where $f(\lambda)=1+\alpha^2 \cos^{2}\lambda$. The 3-sphere ($\ref{HorizonUndis}$) is a deformed or distorted one as compared to ($\ref{Horizonnorotation}$). For $\alpha=0$ the horizon is a round sphere $S^3$. In the presence of rotation, this sphere is
distorted. The shape of the horizon surface depends on the value of the parameter $\alpha$. However, to avoid confusion with the case when this surface is distorted due to the presence of external sources, in this paper, the word distorted corresponds only to the case when at least some of $a_n$ or $b_n$'s are non-zero. For Myers-Perry black hole when, $a_n=b_n=0$, for all values of $n$,  the 3-volume of the horizon surface is 
\begin{equation}
	{\cal A}_H=16\pi^2 \sigma^{3/2} (1+\alpha^2)=2\pi^2\frac{r_0^3}{\sqrt{1+\alpha^2}},
\end{equation}
where $r_0^2=8M/(3\pi)$. 
Using $(\ref{metric_H})$, we can obtain the horizon area for the distorted case
\begin{eqnarray}\label{Area_dist}
	{\cal A}_H &=& \int_H \sqrt{\det{g_H}}d\phi\, d\psi\, dy \nonumber \\[2mm]
	&=& 16\pi^2\sigma^{\frac{3}{2}}(1 + \alpha^2)\exp{\left[-\frac{3}{2}\sum^\infty_{n=0}(a_{2n} + b_{2n}) - \frac{1}{2}\sum^\infty_{n=0}(b_{2n+1} - a_{2n+1})\right]}.
\end{eqnarray}
For the horizon to exist $\sigma>0$. 
If no external matter fields are present, i.e. $a_n=b_n=0$ the expression reduces to the horizon area of the asymptotically flat Myers-Perry black hole.

In order to study the effect of distortions on the black hole, we consider the simplest cases of distortion. Distortions are parametrized by the multiple moments $a_{n}$ or $b_{n}$. The multiple moments decrease in their strength by increasing the order of the multiple moments. The simplest case corresponds to the monopole distortions $a_0$ and $b_0$. However, the pure monopole distortions create trivial deformations of the black hole and its horizon. Thus, we consider monopole-dipole and monopole-quadrupole multiple moments to separate the even and odd multiple moments. 

%In order to study the effect of distortions on the black hole, we consider the simplest cases, i.e., monopole distortions $a_{n}=b_{n}=0$ for $n>0$, dipole distortion, etc. Taking $\hat{W}$ as a dipole distortion and $\hat{U}$ as monopole distortion, and using ($\ref{conicalmp}$) we derive  
%\be
%\hat{U}=b_{0}+3b_{1},~~~ \hat{W}=b_{0}+b_{1}xy~. \label{dipolemono1}
%\ee
%We can also consider $\hat{U}$ as a dipole distortion and $\hat{W}$ as monopole distortion:
%\be
%\hat{W}=a_{0}-3a_{1},~~~ \hat{U}=a_{0}+a_{1}xy~.  \label{dipolemono2}
%\ee
%\noindent 
In considering multiple moments, we impose the no-conical singularity conditions (\ref{conicalmp}).
In the case of the distorted five-dimensional Schwarzchild black hole, i.e., when the rotation parameter vanishes, exchange between these two cases (whether $\hat{U}$ is dipole or $\hat{W}$ is dipole, and the other is monopole) corresponds only to the switch of the axes. Since the black hole is rotating only with respect to one of the axes, this is not true anymore. In what follows, we need to investigate both cases. 

\subsection{Deformations of the Horizon Surface}
Because of the presence of the distortion fields, the shape of the horizon surface changes. Namely, the horizon surface is a deformed $S^{3}$ sphere. To study this deformation, we use the isometric embedding of the horizon sections into the 3-dimensional flat space. Here, the horizon surface is a  3-dimensional one. Such a surface possesses more than one curvature invariant, and its isometric local embedding generally requires $3(3+1)/2=6$ dimensional flat space. However, the horizon surface admits a group of isometries, and one can analyze its geometry by studying the geometry of the sections of the isometry orbits. An isometric embedding ensures that distances and angles measured on the surface remain unchanged when it is placed in a higher-dimensional Euclidean or pseudo-Riemannian space. In general relativity, isometric embeddings are particularly useful for studying black hole horizons, cosmological models, and the global properties of spacetime. 

The geometrical properties of the horizon surface of five-dimensional undistorted rotating black holes and black rings have been studied in \cite{Frolov:2006pu}. Here, we study the distorted case. We shall consider the embedding of the $(\psi, \theta)$ 2-dimensional section defined by $\phi= const.$, and $(\phi, \theta)$ 2-dimensional section defined by $\psi=const$. The section $(\psi, \phi)$ defined by $\theta=const.$ represents a 2-dimensional torus whose radii are defined by the distortion fields. We do not consider the embeddings of this section. 

We can visualize the effect of the distortion fields on the horizon surface by considering isometric embeddings of its 2-dimensional sections of the horizon surface into a flat 3-dimensional space with the following metric in the cylindrical coordinates:
\be
dl^{2}=\epsilon dZ^{2}+d\rho^{2}+\rho^{2}d\varphi^{2},
\ee
where $\epsilon=1$ corresponds to Euclidean space, and $\epsilon=-1$ corresponds to pseudo-Euclidean space. 
The horizon surface metric in the $(\psi,\theta)$ section is expressed as
\begin{eqnarray}
	ds^2_{H\psi\theta} &=& \frac{4\sigma(1+\alpha^2)^2}{1 + {\hat a}^2(\theta)\cos^2(\frac{\theta}{2})}\sin^2(\frac{\theta}{2})e^{-2{\widehat W}(\theta)}d\psi^2\nonumber\\
	&+& \sigma\left[1 + {\hat a}^2(\theta)\cos^2(\frac{\theta}{2})\right]e^{2({\hat\gamma}(\theta)-{\widehat W}(\theta))} d\theta^2.\label{metricsection1}
\end{eqnarray}
The metric can be parametrized in the cylindrical coordinates as follows:
\be
Z=Z(\theta),~~~\rho=\rho(\theta). 
\ee
The geometry induced on the section is given by
\be\label{Embed}
dl^{2}=(\epsilon Z_{,\theta}^{2}+ \rho_{,\theta}^{2}) d\theta^{2}+\rho^{2}d\varphi^{2}. 
\ee
Matching the metrics ($\ref{metricsection1}$) and ($\ref{Embed}$), we derive the embedding map
\ba
&&\hspace{-1cm}\psi=\varphi~~~\rho(\theta)=\frac{2\sqrt{\sigma}(1+\alpha^{2})}{[1 + {\hat a}^2(\theta)\cos^2(\frac{\theta}{2})]^{\frac{1}{2}}}\sin(\frac{\theta}{2}) e^{-{\widehat W}(\theta)},\nonumber\\
&&\hspace{-1cm}Z(\theta)=\int_{\pi}^{\theta}Z_{,\theta'}d\theta',~\\
&&Z_{,\theta}=\left[\epsilon\left(\sigma\left[1 + {\hat a}^2(\theta)\cos^2(\frac{\theta}{2})\right]e^{2({\hat\gamma}(\theta)-{\widehat W}(\theta))} -\rho_{,\theta}^2\right)\right]^{\frac{1}{2}}=[\epsilon\mathcal{F_\psi}]^{\frac{1}{2}}\label{EmbedMap1}
\ea
The metric of the section $(\phi,\theta)$ reads
\begin{eqnarray}
	ds^2_{H\phi\theta} &=&4\sigma \cos^2(\frac{\theta}{2}) e^{-2{\widehat U}(\theta)}d\phi^2 + \sigma\left[1 + {\hat a}^2(\theta)\cos^2(\frac{\theta}{2})\right]e^{2({\hat\gamma}(\theta)-{\widehat W}(\theta))} d\theta^2.\label{metricsection2}
\end{eqnarray}
Thus, the embedding map reads
\ba
&&\hspace{-1cm} \phi=\varphi~~~~\rho(\theta)=2\sqrt{\sigma} \cos(\frac{\theta}{2}) e^{-{\widehat U}(\theta)},\nonumber\\
&&\hspace{-1cm}Z(\theta)=\int_{0}^{\theta}Z_{,\theta'}d\theta',\\
&&Z_{,\theta}=\left[\epsilon\left(\sigma\left[1 + {\hat a}^2(\theta)\cos^2(\frac{\theta}{2})\right]e^{2({\hat\gamma}(\theta)-{\widehat W}(\theta))} -\rho_{,\theta}^2\right)\right]^{\frac{1}{2}}=[\epsilon\mathcal{F_\phi}]^{\frac{1}{2}}\label{EmbedMap2}. 
\ea
%% Embedding for Undistorted Case

For some value of $\alpha$ the embedding of the horizon surface into the Euclidean space, $\mathbb{E}^{3}$ becomes impossible. 
In all the embedding figures presented in this paper, the dotted line corresponds to the region of the horizon surface, where the isometric embedding of the horizon surface into the Euclidean space is not possible. This situation arises when the quantity beneath the square root in (\ref{EmbedMap1}) or (\ref{EmbedMap2}) is negative for the case $\epsilon = 1$, and we need to consider $\epsilon=-1$ corresponding to pseudo-Euclidean embedding. 

The simplest non-trivial type of distortion is described by setting $b_1\neq 0$ or $a_1\neq 0$ and requiring the other external matter parameters to vanish. If we assume that the multiple moments are due to the external sources and matter, we expect the higher-order multiple moments to have a smaller magnitude. Therefore, dipole or quadrupole are the dominant terms. 

To analyze the effect of various multiple moments of the simplest kind, we can consider four different cases: Case I is the monopole-dipole distortion of $\hat{W}$. This corresponds to $b_0=-3b_1\neq 0$, $a_0=a_1=b_2=a_2=0$. In case II, we have monopole-dipole distortion of mixed type $b_0=b_1\neq 0$ and $a_0=a_1=-2b_0$. In case III: monopole-dipole distortion of $\hat{U}$, where  $a_0=3a_1\neq 0$, $b_0=b_1=b_2=a_2=0$. In case IV, we consider the quadrupole-quadrupole case of $\hat{W}=\hat{U}$. All these cases are constructed such that the no-conical singularity condition is respected. 

\begin{figure}[htp]
	\setlength{\tabcolsep}{ 0 pt }{\scriptsize\tt
		\begin{tabular}{ cccc }
			\includegraphics[width=15 cm]{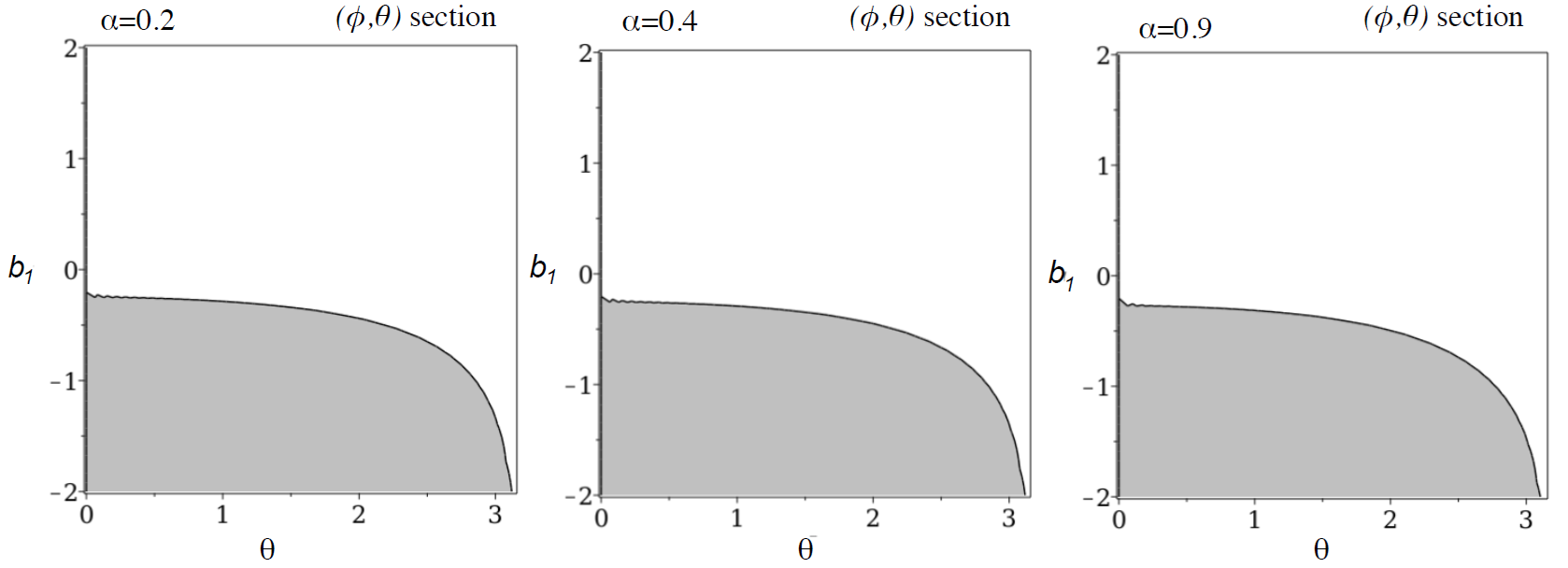} \\
			\includegraphics[width=15 cm]{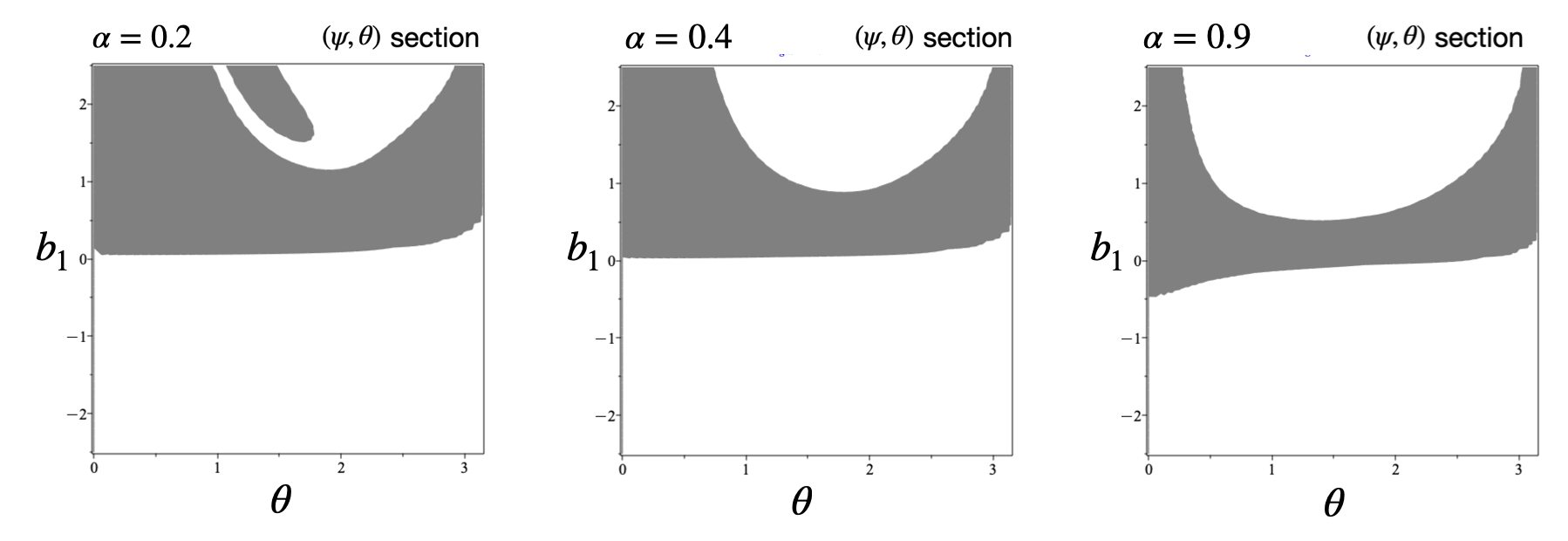} \\
	\end{tabular}}
	\caption{\footnotesize{Contour plots of functions $\mathcal{F_\phi}$ and $\mathcal{F_\psi}$ , where the first and second row respectively show function $\mathcal{F_\phi}$ and $\mathcal{F_\psi}$. Gray areas correspond to the negative values. This corresponds to the monopole-dipole distortion of $\hat{W}$ or case I, where, $b_0=-3b_1$, $a_0=a_1=0$}}
	\label{Contour1}	
\end{figure}

\begin{figure}[htp]
	\setlength{\tabcolsep}{ 0 pt }{\scriptsize\tt
		\begin{tabular}{ cccc }
			\includegraphics[width=15 cm]{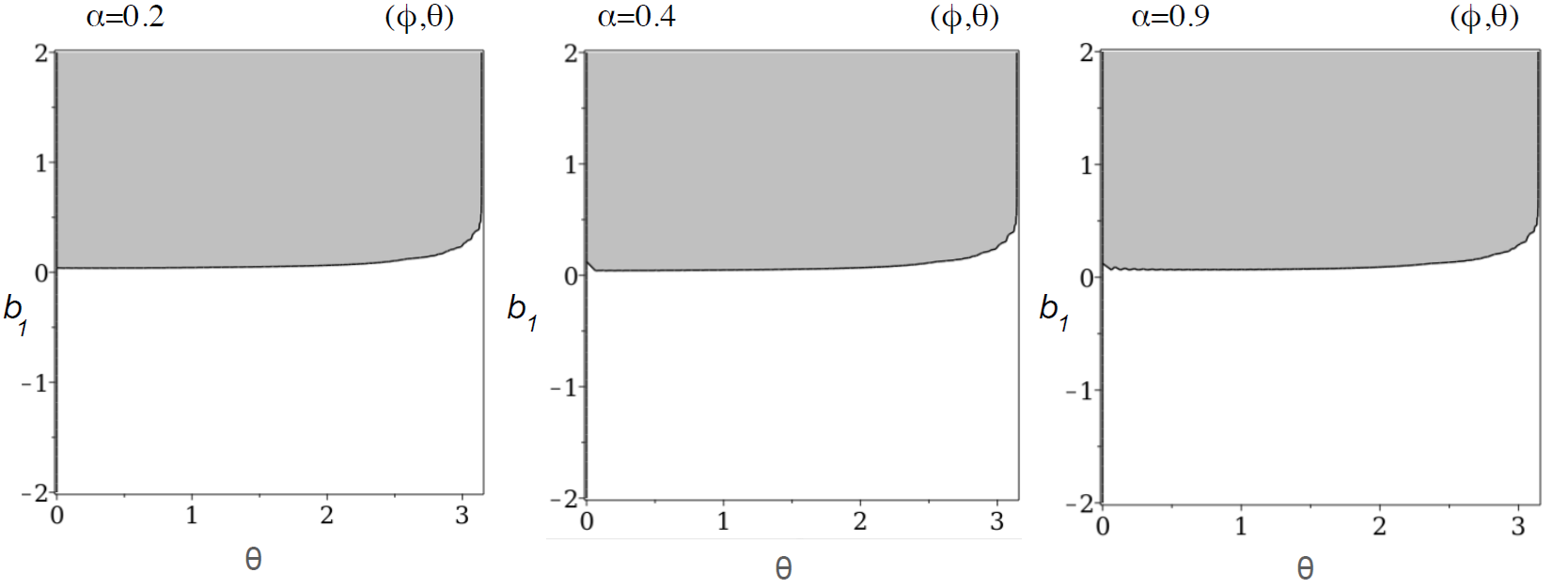} \\
			\includegraphics[width=15 cm]{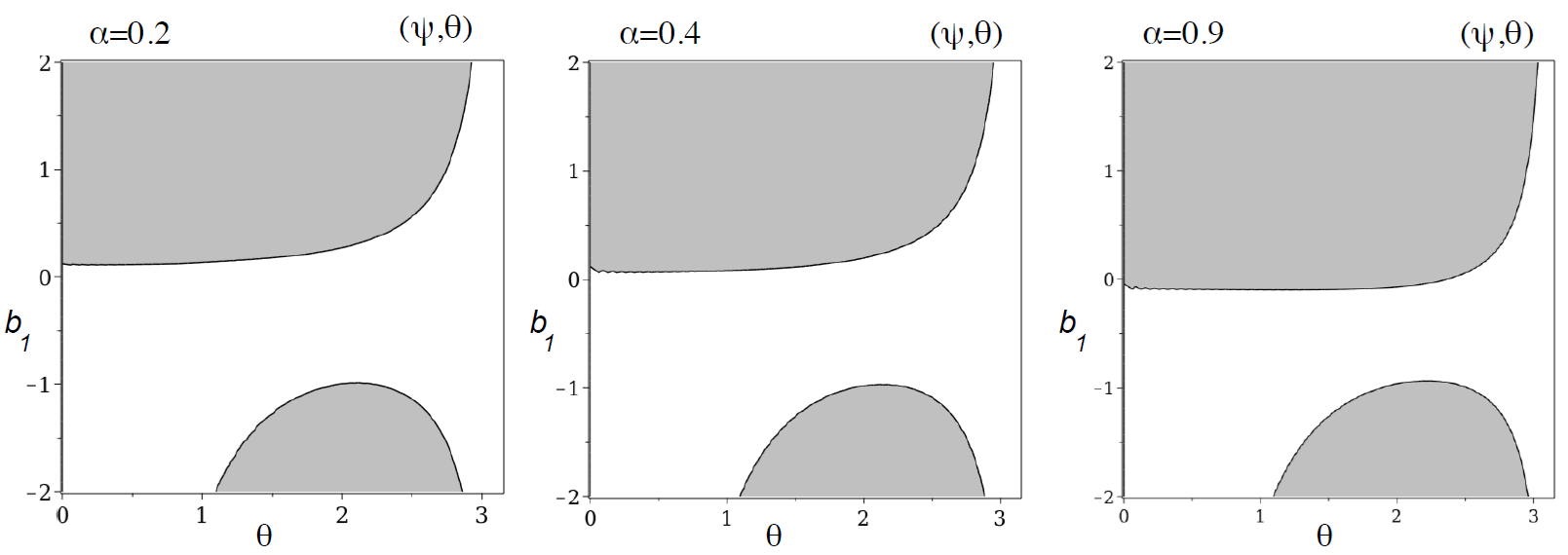} \\
	\end{tabular}}
	\caption{\footnotesize{Contour plots of functions $\mathcal{F_\phi}$ and $\mathcal{F_\psi}$ , where the first and second row respectively show function $\mathcal{F_\phi}$ and $\mathcal{F_\psi}$. Gray areas correspond to the negative values. This corresponds to the monopole-dipole distortion of mixed type $\hat{W}$ and $\hat{U}$ or case II, where, $b_0=b_1$, $a_0=a_1=-2b_{1}$}}
	\label{Contour2}
\end{figure}

\begin{figure}[htp]
	\setlength{\tabcolsep}{ 0 pt }{\scriptsize\tt
		\begin{tabular}{ cccc }
			\includegraphics[width=15 cm]{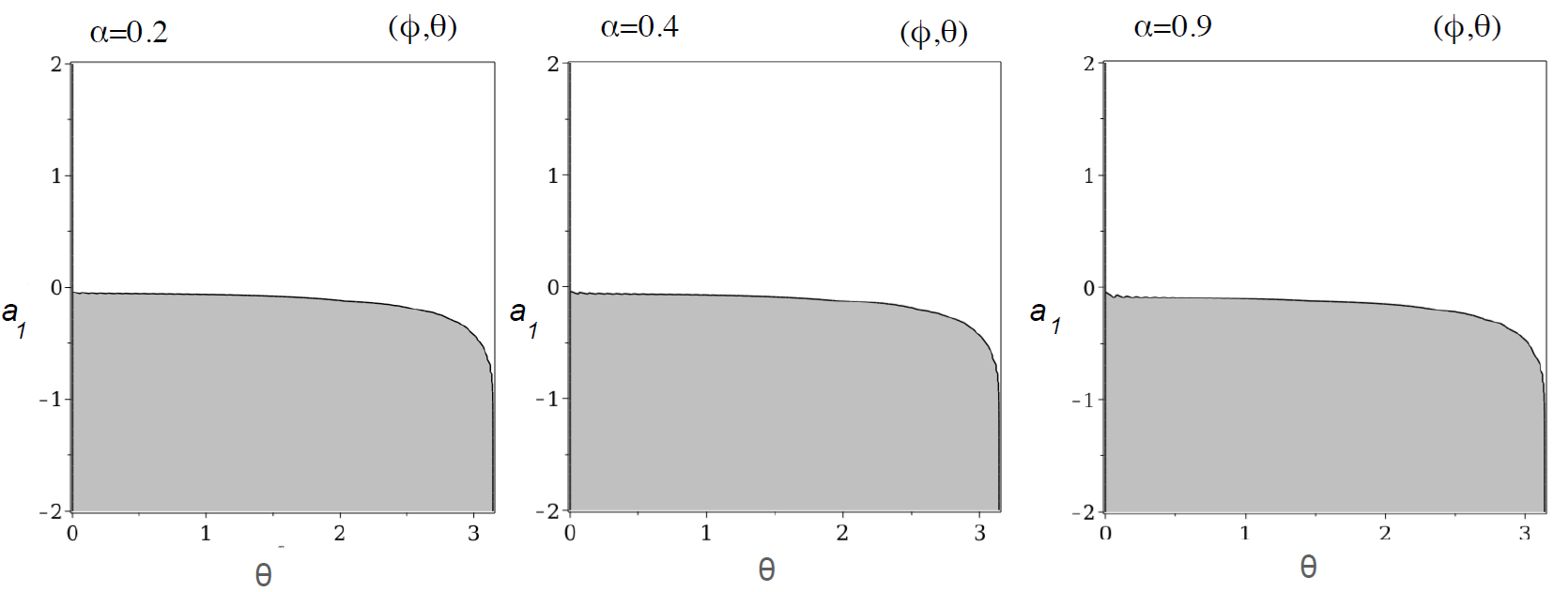} \\
			\includegraphics[width=15 cm]{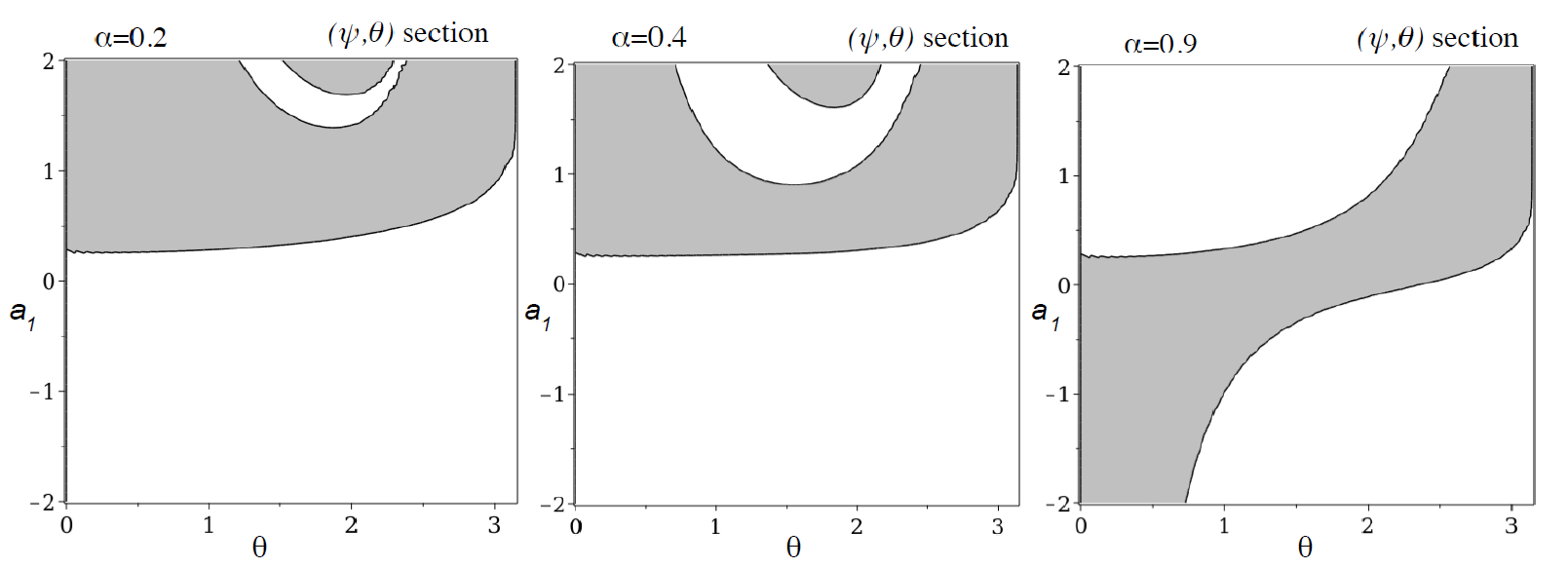} \\
	\end{tabular}}
	\caption{\footnotesize{Contour plots of functions $\mathcal{F_\phi}$ and $\mathcal{F_\psi}$ , where the first and second row respectively show function $\mathcal{F_\phi}$ and $\mathcal{F_\psi}$. Gray areas correspond to the negative values. This corresponds to the monopole-dipole distortion of $\hat{U}$ or case III, where, $b_0=b_1=0$, $a_0=3a_1$}}
	\label{Contour3}
\end{figure}

\begin{figure}[htp]
	\setlength{\tabcolsep}{ 0 pt }{\scriptsize\tt
		\begin{tabular}{ cccc }
			\includegraphics[width=15 cm]{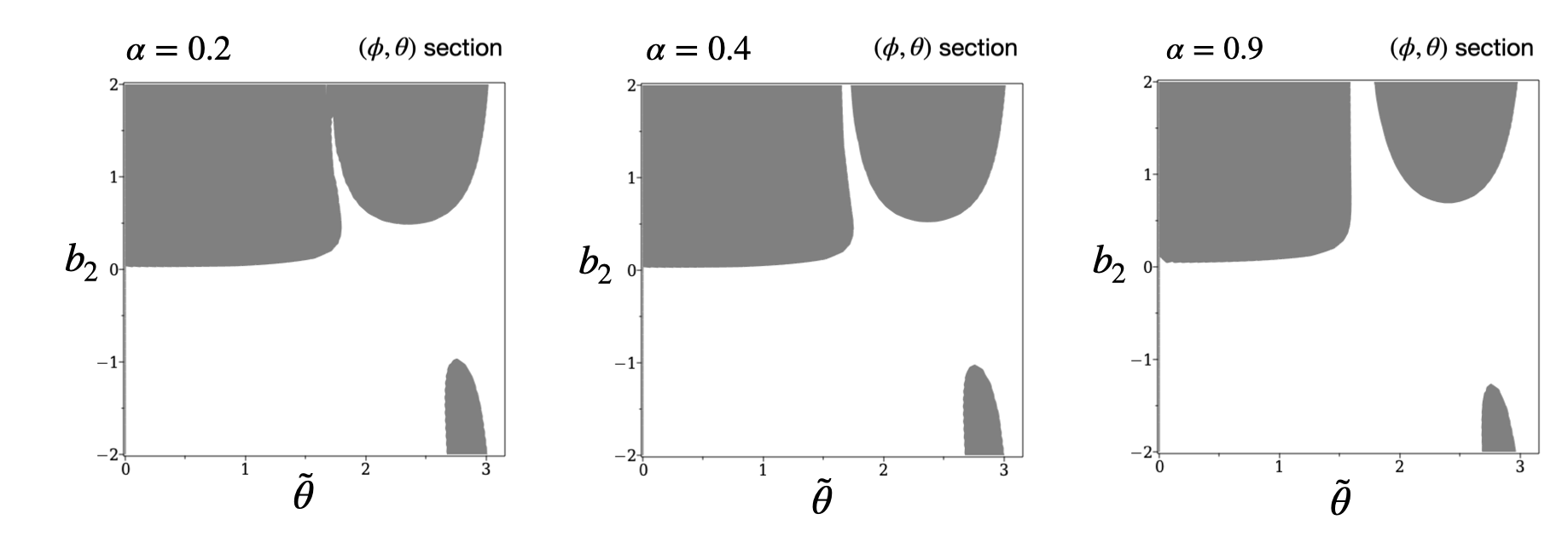} \\
			\includegraphics[width=15 cm]{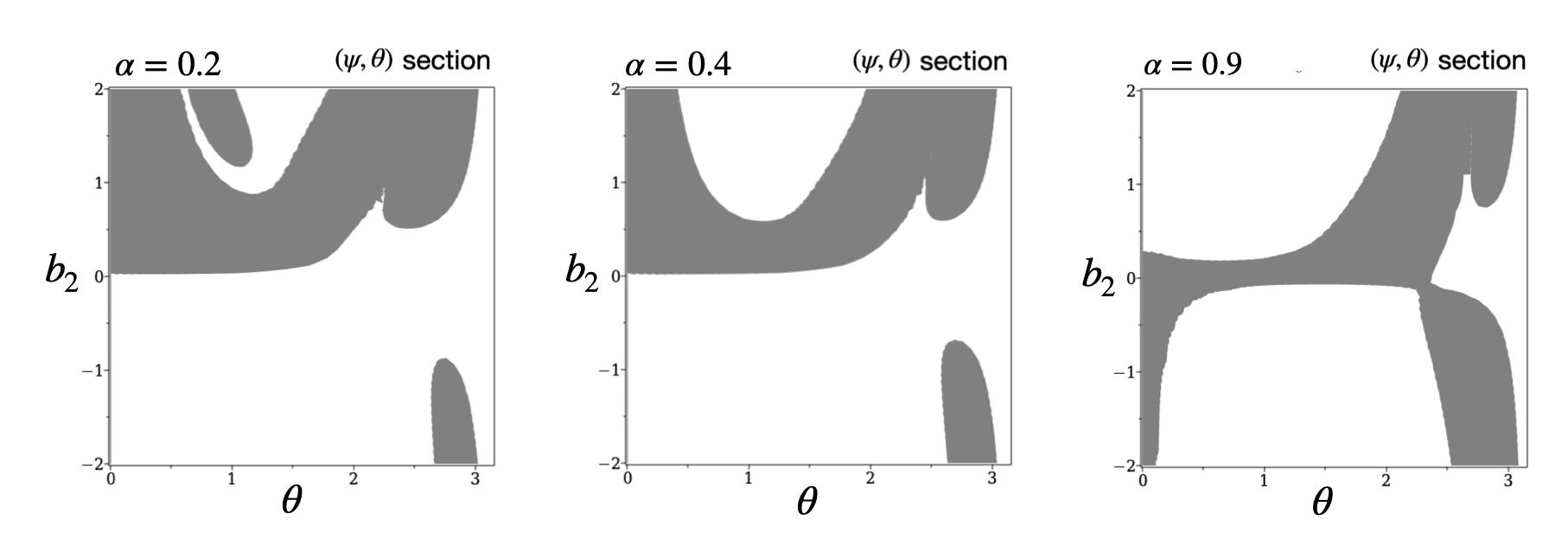} \\
	\end{tabular}}
	\caption{\footnotesize{Contour plots of functions $\mathcal{F_\phi}$ and $\mathcal{F_\psi}$ , where the first and second row respectively show function $\mathcal{F_\phi}$ and $\mathcal{F_\psi}$. Gray areas correspond to the negative values. Quadrupole case IV is considered. $b_0=a_0=b_2=a_2$, $a_1=b_1=0$. }}
	\label{Contour1_posV2}	
\end{figure}

Figures \ref{Contour1}-\ref{Contour1_posV2} show the possibility or lack of the embedding of 2-dimensional ($\phi,\theta)$ and  ($\psi,\theta)$ sections of the horizon surface into the Euclidean 3-dimensional space. In each case, we fix $\alpha$ and illustrate for different values of multiple moments whether $\mathcal{F_\psi}$ or $\mathcal{F_\phi}$ is positive for all angles. In the angles with negative values of $\mathcal{F_\psi}$ or $\mathcal{F_\phi}$, we use pseudo-Euclidian embedding with $\epsilon=-1$.  In the gray (shaded) parameter space the embedding in Euclidean space is not possible, and one should use the pseduo-Euclidean embedding. In the white regions of the parameter space, the embedding in the Euclidean space is possible.

The first row of Figure \ref{Contour1}-\ref{Contour1_posV2}, shows the region where embedding is not possible for in the parameter space of $b_1$ and $\theta$ for $\alpha=0.2,0.4,0.9$ for the ($\phi,\theta$). The second rows of figures \ref{Contour1}-\ref{Contour1_posV2} show the embedding of the $(\psi,\theta)$ section of the horizon surface. According to figures \ref{Contour1} and \ref{Contour3} for monopole-dipole distortion, the embedding in the Euclidean space is possible for all positive values of $b_1$ or $a_1$. However, there is minimal dependence on the value of $\alpha$.


\begin{figure}[htp]
	\setlength{\tabcolsep}{ 0 pt }{\scriptsize\tt
		\begin{tabular}{ cccc }
			\includegraphics[width=15 cm]{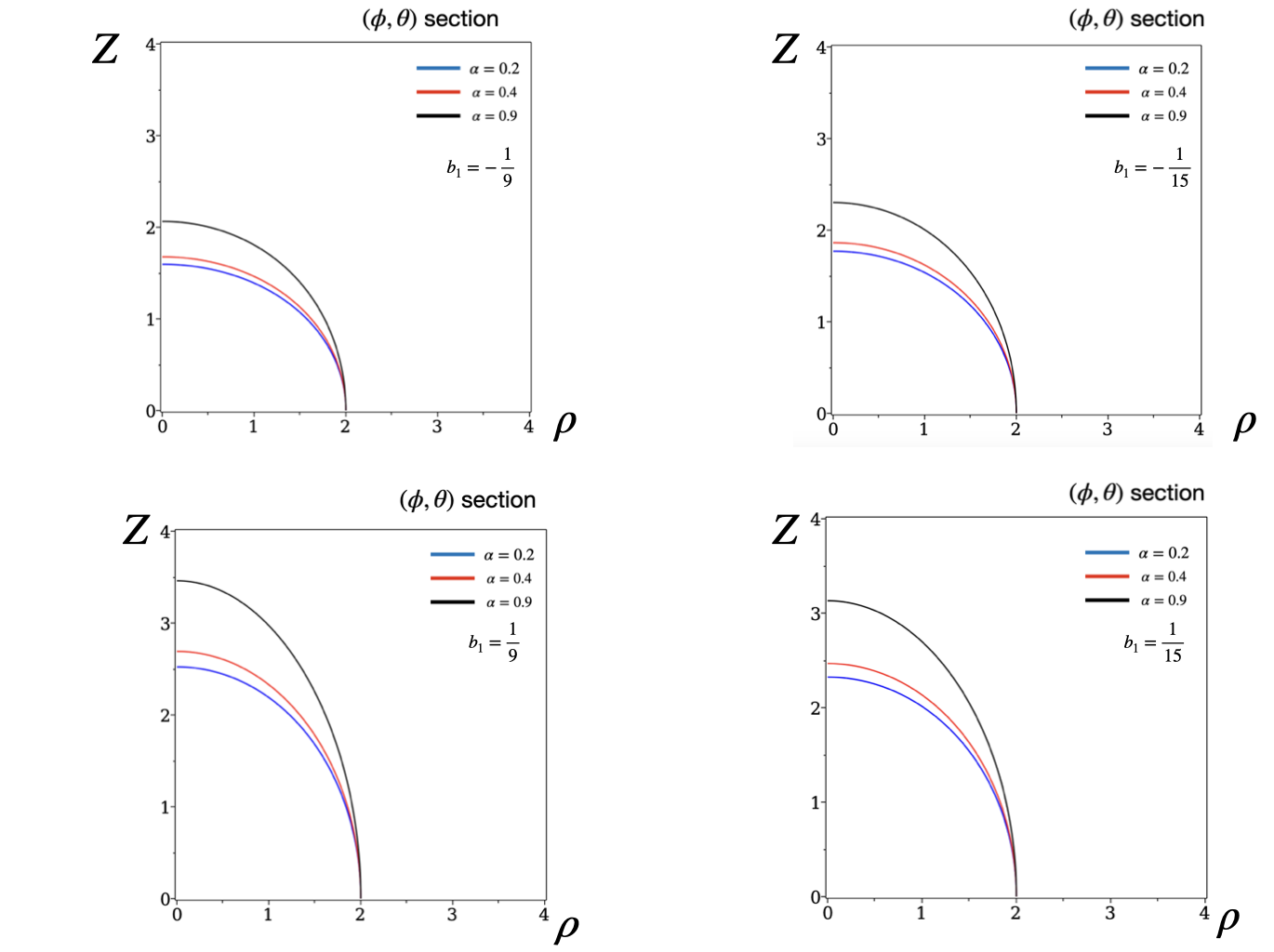} \\
	\end{tabular}}
	\caption{\footnotesize{Rotational curves illustrating embeddings of the $(\phi,\theta)$ section of the horizon surface for different values of $\alpha$. Monopole-dipole case of $b_1$ is considered, $b_0=-3b_1$, $a_0=a_1=0$ (Case I).}}
	\label{EmbedPhi1}
\end{figure}

\begin{figure}[htp]
	\setlength{\tabcolsep}{ 0 pt }{\scriptsize\tt
		\begin{tabular}{ cccc }
			\includegraphics[width=15 cm]{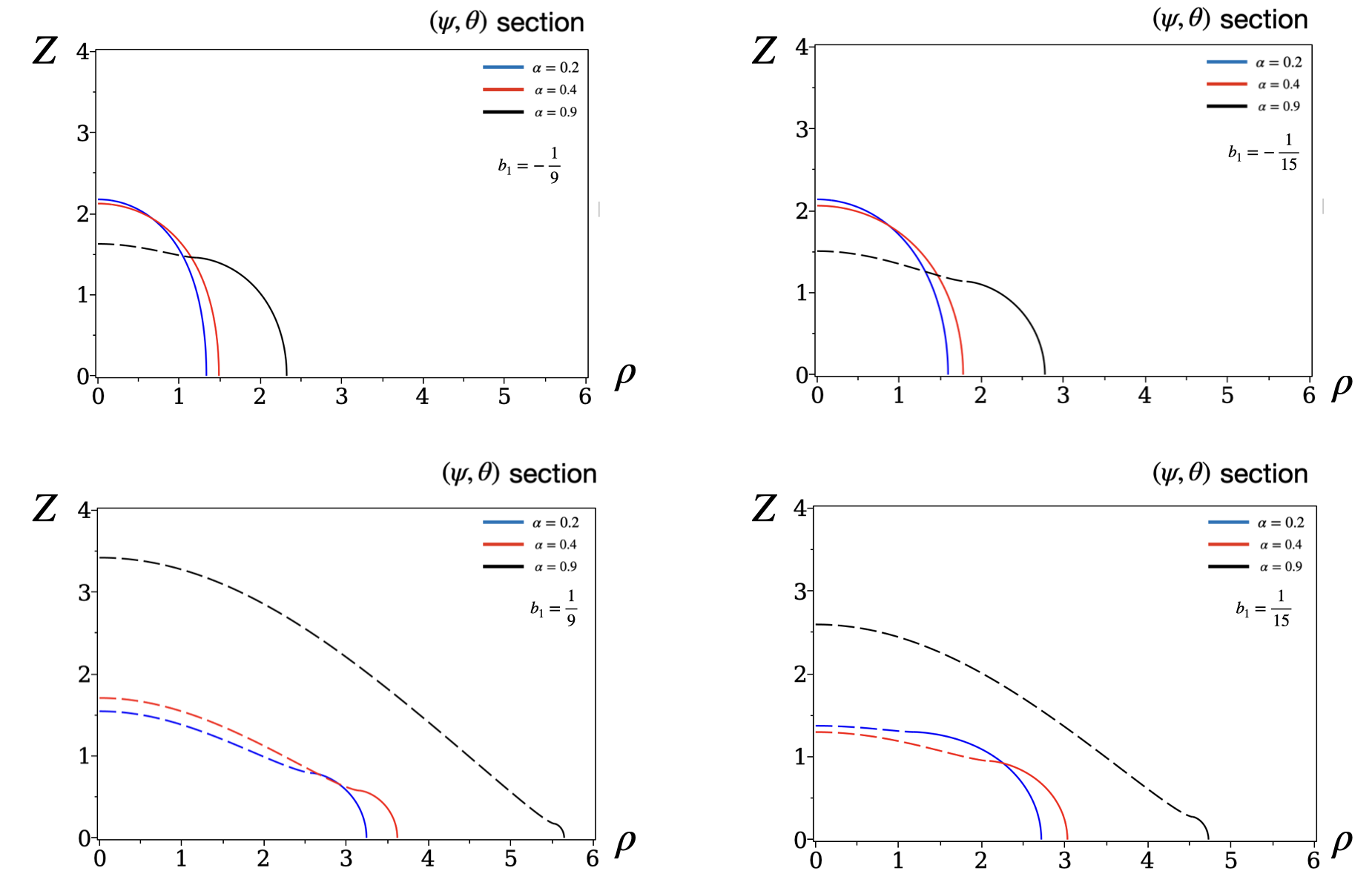} \\
	\end{tabular}}
	\caption{\footnotesize{Rotational curves illustrating embeddings of the $(\psi,\theta)$ section of the horizon surface for different values of $\alpha$. Monopole-dipole case of $b_1$ is considered, $b_0=-3b_1$, $a_0=a_1=0$, (case I). }}
	\label{EmbedPsi1}
\end{figure}

\begin{figure}[htp]
	\setlength{\tabcolsep}{ 0 pt }{\scriptsize\tt
		\begin{tabular}{ cccc }
			\includegraphics[width=15 cm]{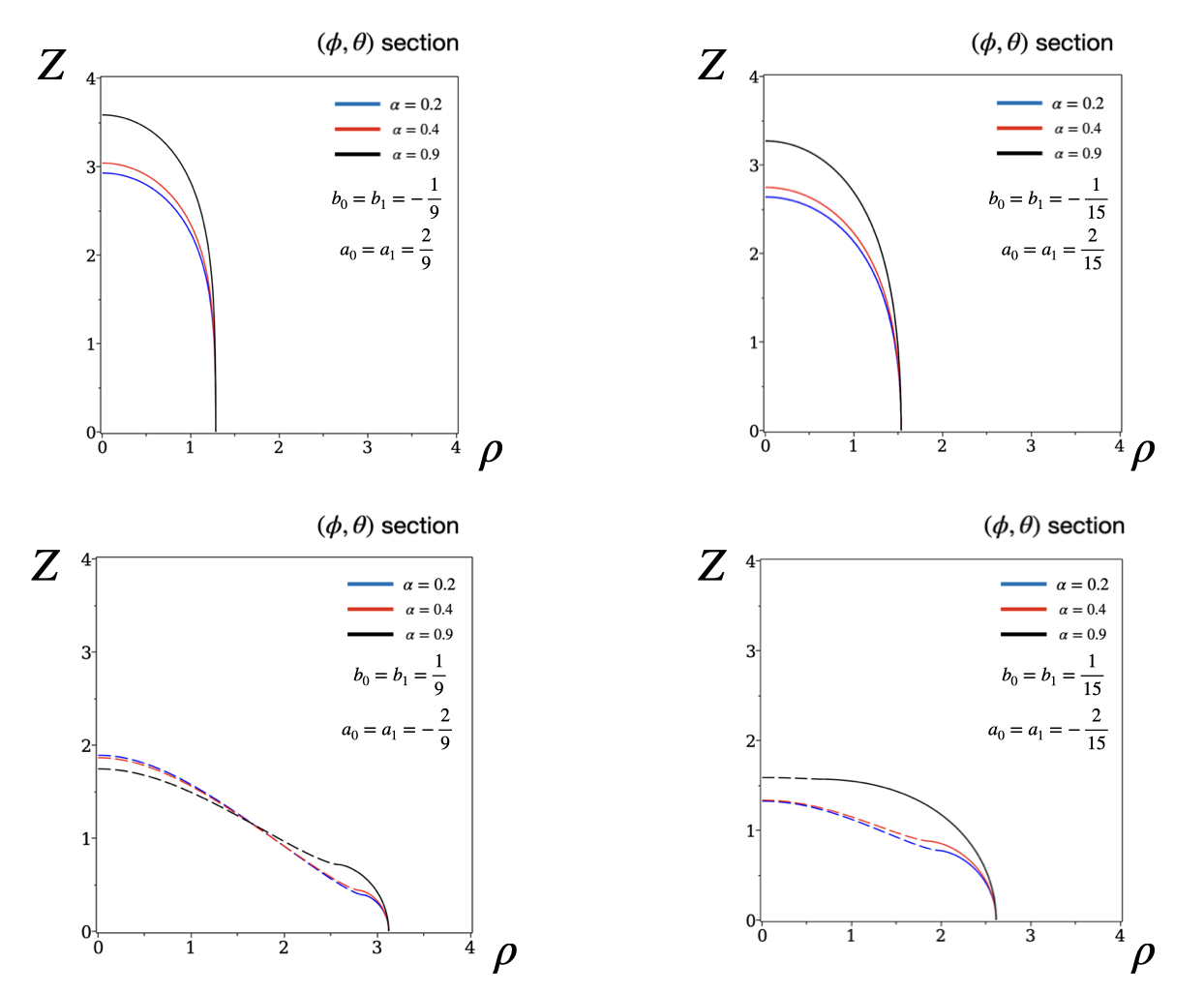} \\
	\end{tabular}}
	\caption{\footnotesize{Rotational curves illustrating embeddings of the $(\phi,\theta)$ section of the horizon surface for different values of $\alpha$.We consider the monopole-dipole case $b_0=b_1$, $a_0=a_1=-2b_0$ (case II).}}\label{EmbedPhi1V2}
\end{figure}

\begin{figure}[htp]
	\setlength{\tabcolsep}{ 0 pt }{\scriptsize\tt
		\begin{tabular}{ cccc }
			\includegraphics[width=15 cm]{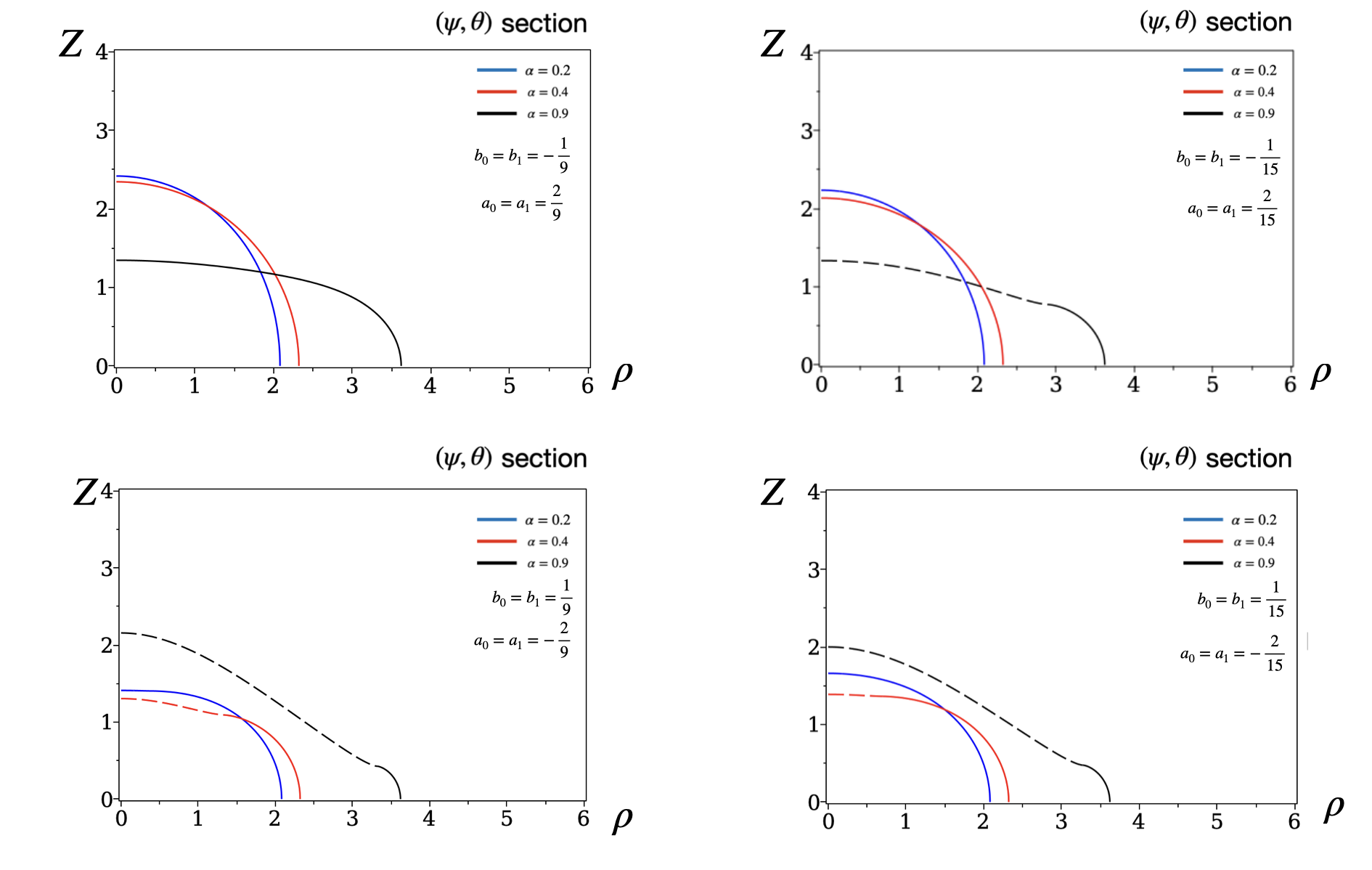} \\
	\end{tabular}}
	\caption{\footnotesize{Rotational curves illustrating embeddings of the $(\psi,\theta)$ section of the horizon surface for different values of $\alpha$.We consider the monopole-dipole case $b_0=b_1$, $a_0=a_1=-2b_0$ (case II).}}\label{EmbedPsiV2}
\end{figure}

\begin{figure}[htp]
	\setlength{\tabcolsep}{ 0 pt }{\scriptsize\tt
		\begin{tabular}{ cccc }
			\includegraphics[width=15 cm]{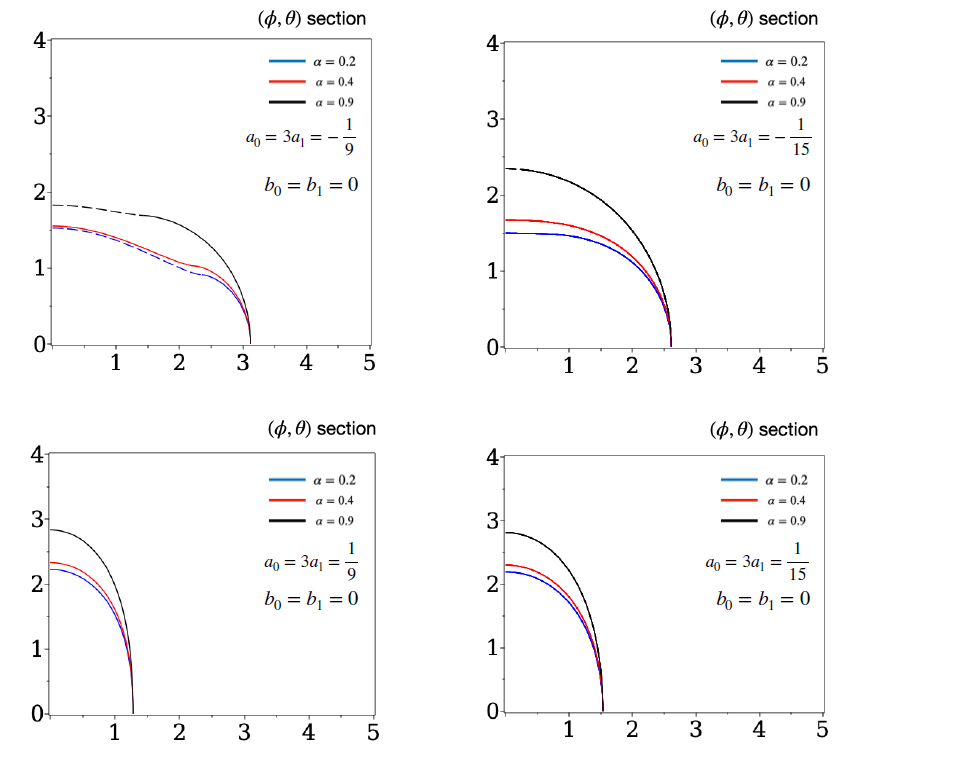} \\
	\end{tabular}}
	\caption{\footnotesize{Rotational curves illustrating embeddings of the $(\phi,\theta)$ section of the horizon surface for different values of $\alpha$. We consider monopole-dipole case where $ a_0 = 3a_1$ and $b_0 = b_1 = b_2 = a_2 = 0$ (case III). }}
	\label{phi-th3}
\end{figure}

\begin{figure}[htp]
	\setlength{\tabcolsep}{ 0 pt }{\scriptsize\tt
		\begin{tabular}{ cccc }
			\includegraphics[width=15 cm]{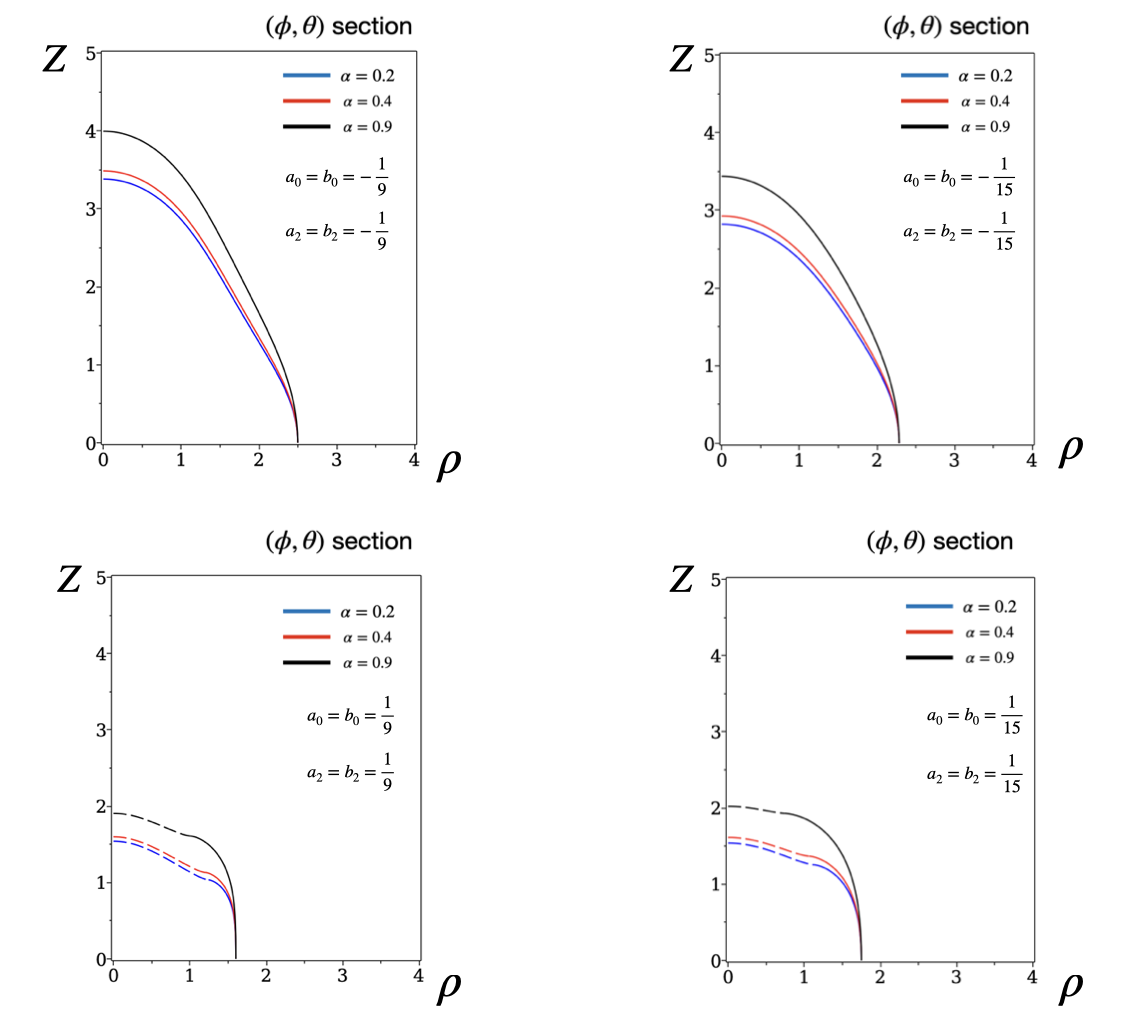} \\
	\end{tabular}}
	\caption{\footnotesize{Rotational curves illustrating embeddings of the $(\phi,\theta)$ section of the horizon surface for different values of $\alpha$.We consider the monopole-quadruple case $a_1=b_1=0$, $a_0=b_0=a_2=b_2$ (Case IV).}}
	\label{EmbedPhi1V3}
\end{figure}

\begin{figure}[htp]
	\setlength{\tabcolsep}{ 0 pt }{\scriptsize\tt
		\begin{tabular}{ cccc }
			\includegraphics[width=15 cm]{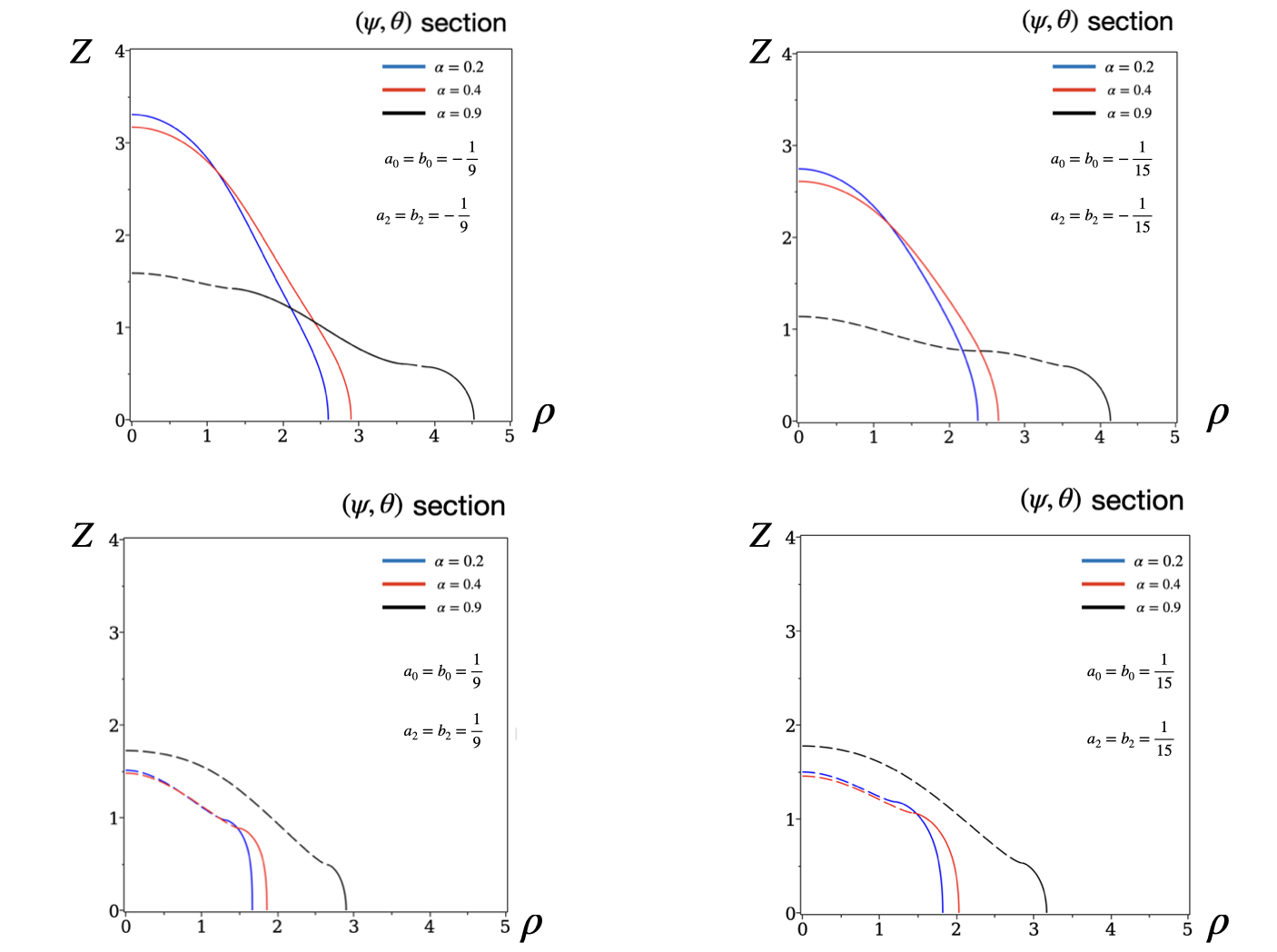} \\
	\end{tabular}}
	\caption{\footnotesize{Rotational curves illustrating embeddings of the $(\psi,\theta)$ section of the horizon surface for different values of $\alpha$. We consider the monopole-quadruple case $a_1=b_1=0$, $a_0=b_0=a_2=b_2$ (Case IV).}}
	\label{EmbedPSiV3}
\end{figure}

In Figs. \ref{EmbedPhi1}-\ref{EmbedPSiV3}, we illustrate the rotational curves for the embeddings of the horizon surface of an undistorted Meyers-Perry black hole. To reconstruct the shape of the 3-dimensional horizon surface, we have to rotate these curves around the axes $\theta=0$ or $\theta=\pi$, which corresponds to the vertical line in all the embeddings. 

For the undistorted black hole and the $(\psi,\theta)$ section of the horizon surface, the larger the value of $\alpha$, the more oblate the horizon surface is. For the undistorted black hole and the $(\phi,\theta)$ section of the horizon surface, the larger the value of $\alpha$, the more prolate the surface of the horizon becomes. 
\begin{figure}[htbp]
	\begin{center}
		\includegraphics[width=16cm]{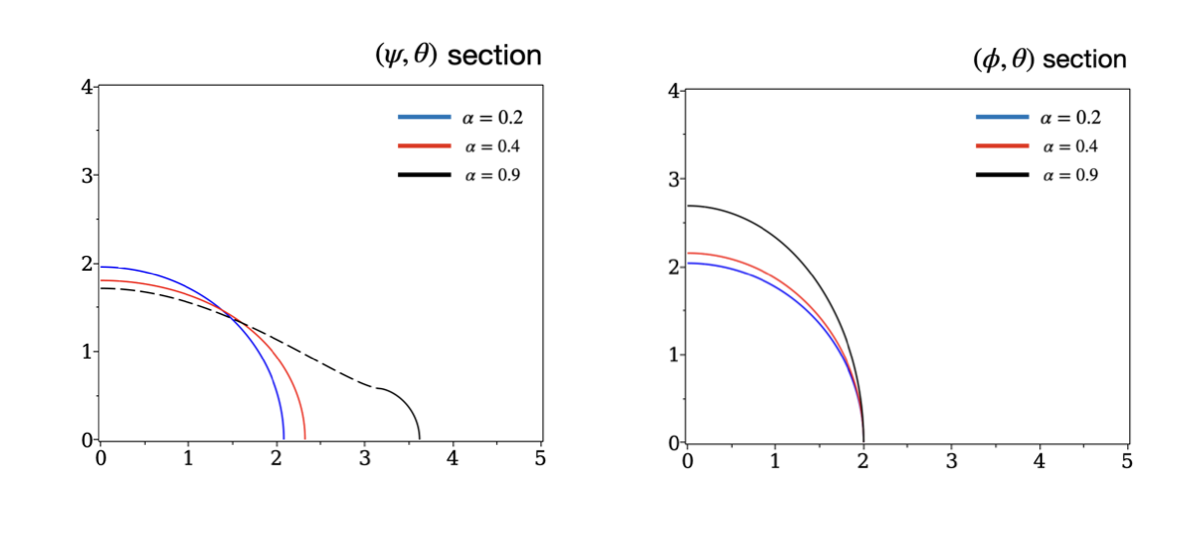}~~
		\caption{
			Rotational curves illustrating undistorted embeddings of the horizon surface for different values of $\alpha$ for $(\psi,\theta)$ and $(\phi,\theta)$ sections. }\label{Fig1}
	\end{center}
\end{figure}

Figures \ref{EmbedPhi1} and \ref{EmbedPSiV3} illustrate the embedding of the 2-dimensional ($\phi,\theta)$ and ($\psi,\theta)$ section of the horizon surface into the Euclidean 3-dimensional space for monopole-dipole case of $b_0=-3b_1$ (case I). For a fixed value of $\alpha$, the larger positive values of $b_1$ or $a_1$ correspond to a more prolate horizon surface.

Figures \ref{EmbedPhi1V2} and \ref{EmbedPsiV2} illustrate the embedding of the 2-dimensional ($\phi,\theta)$ and ($\psi,\theta)$ section of the horizon surface into the Euclidean 3-dimensional space for case II. For a fixed value of $\alpha$, when $|a_1|$ increases, the horizon surface in both sections becomes more prolate. 

%
%
%Figure 1, alphas%
%
%
%\begin{figure}[htbp]
%\begin{center}
%\includegraphics[width=6cm]{fig2apart3.pdf}~~
%\includegraphics[width=6cm]{fig2bpart3.png}~~
%\caption{
	%Rotational curves illustrating undistorted embeddings of the horizon surface for different values of $\alpha$, line 1: $\alpha=1/4$, line 2: $\alpha=0.8$, and line 3: $\alpha=1$; (a) $(\psi,\theta)$ section, (b) $(\phi,\theta)$ section. }\label{Fig1}
%\end{center}
%\end{figure}

Case IV is of special interest as it belongs to a special case of deformations with very interesting property. In this special case, the multiple moments satisfy the condition $a_n+2b_n=0$ for every $n$. In this setup, the angular velocity of the horizon remains unchanged compared to the undistorted case. The ergoregion behaves as in the asymptotically flat undistorted case scenario. The absence of conical singularities requires satisfying the condition 
\begin{equation}
	\sum_{n=0}^{\infty} (-1)^n b_n=0, 
\end{equation}
which matches requiring a Minkowski deformed spacetime free of conical singularities. In other words, ergoregion remains undistorted, and the effect of distortion vanishes in many parts of the spacetime. However, figures (\ref{EmbedPsiV2}) and (\ref{EmbedPhi1V2}) represent that a deformed horizon exists even with a non-deformed ergoregion. 


\begin{figure}[htp]
	\setlength{\tabcolsep}{ 0 pt }{\scriptsize\tt
		\begin{tabular}{ cccc }
			\includegraphics[width=15 cm]{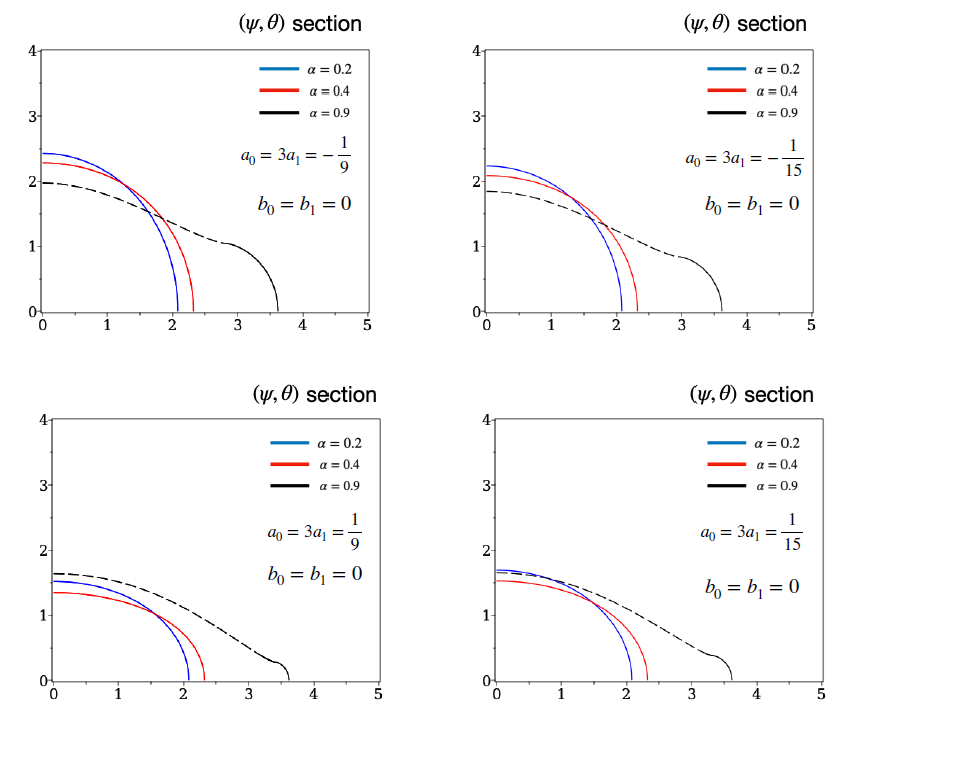} \\
	\end{tabular}}
	\caption{\footnotesize{Rotational curves illustrating embeddings of the $(\psi,\theta)$ section of the horizon surface for different values of $\alpha$. We consider monopole-dipole case where $ a_0 = 3a_1$ and $b_0 = b_1 = b_2 = a_2 = 0$ (case III). }}
	\label{}
\end{figure}


\subsection{Curvature of the Sections of the horizon surface}
The conditions for the isometric embedding of 2D manifolds in 3-dimensional Euclidean space are well-examined. Any compact surface embedded isometrically in $\mathbb{E}^{3}$  has at least one point of positive Gauss curvature. Any 2D compact surface with positive Gauss curvature is always isometrically embeddable in $\mathbb{E}^{3}$, and this embedding is unique up to rigid rotations \cite{Berger}. For a 2-dimensional axisymmetric metric, if the Gauss curvature is negative at the fixed points of the rotation group, it is impossible to isometrically embed a region containing such a fixed point in $\mathbb{E}^{3}$. Such surfaces can be globally embedded in $\mathbb{E}^{4}$. Gauss curvature, $K$ is related to the Ricci scalar $\mathcal{R}$, $K=\mathcal{R}/2$. 

Given the two-dimensional metrics ($\ref{metricsection1}$) and ($\ref{metricsection2}$) of the $(\phi,\theta)$ and $(\psi,\theta)$ horizon subsections, we can calculate their Ricci scalars. Gauss curvature, $K$ is related to the Ricci scalar $\mathcal{R}$, $K=\mathcal{R}/2$. To see whether the isometric embeddings of these 2-dimensional axisymmetric metrics ($\ref{metricsection1}$) and ($\ref{metricsection2}$) into the $\mathbb{E}^{3}$ is possible, we focus on the Ricci scalar of metrics calculated at $\theta=0$ and $\theta=\pi$. The Ricci scalar of metric  ($\ref{metricsection1}$) calculated at $\theta=0$ and $\theta=\pi$ is given by  is given by 
\ba
&&\mathcal{R}_{\psi\theta}|_{\theta=0}=\frac{6e^{a_{0}+b_{0}-3a_{1}-b_{1}+a_{2}+b_{2}}}{\sigma(1+\alpha^{2})^{2}}
\left[\alpha^{2}(a_{1}+2a_{2}+\frac{4}{3}(b_{1}+2b_{2})-\frac{1}{4})\right.\nonumber\\
&&\hspace{5cm}-\frac{1}{3}(a_{1}+2a_{2}+4b_{1}+8b_{2})+\frac{1}{12}\left.\right],\\
&&\mathcal{R}_{\psi\theta}|_{\theta=\pi}=\frac{e^{a_{0}+b_{0}+a_{1}+3b_{1}+a_{2}+b_{2}}}{2\sigma}\left[\alpha^{2} e^{4a_{1}+8b_{1}}+4b_{1}-8b_{2}+1\right].
\ea
These expressions are calculated for general multiple moments $a_{n}$, $b 
_{n}$, $n<3$. 
Ricci scalar of the metric ($\ref{metricsection2}$) calculated at $\theta=0$ and $\theta=\pi$ is given by 
\ba
&&\mathcal{R}_{\phi\theta}|_{\theta=0}=\frac{e^{a_{0}+b_{0}-3a_{1}-b_{1}+a_{2}+b_{2}}}{2\sigma(1+\alpha^{2})}\left[1-4a_{1}-8a_{2}\right],\\
&&\mathcal{R}_{\phi\theta}|_{\theta=\pi}=\frac{e^{a_{0}+b_{0}+a_{1}+3b_{1}+a_{2}+b_{2}}}{2\sigma}\left[\alpha^{2}e^{4a_{1}+8b_{1}}+16a_{1}-32a_{2}+4b_{1}-8b_{2}+1\right].
\ea

\begin{figure}[htp]
	\setlength{\tabcolsep}{ 0 pt }{\scriptsize\tt
		\begin{tabular}{ cccc }
			\includegraphics[width=15 cm]{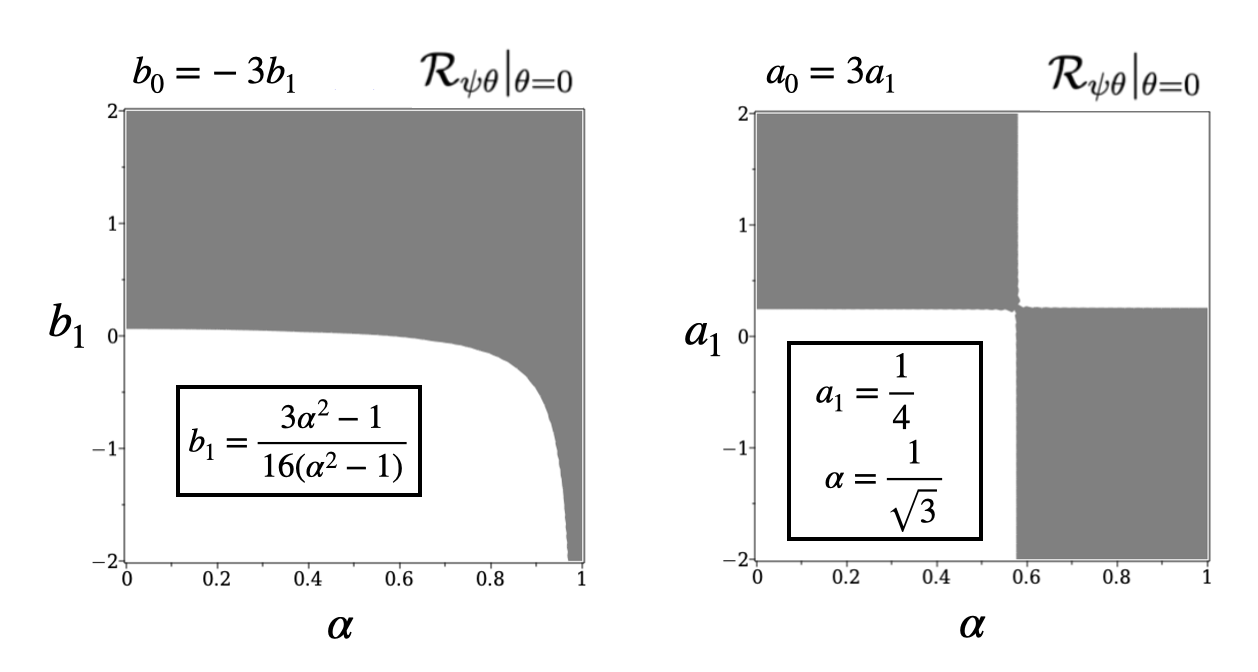} \\
	\end{tabular}}
	\caption{\footnotesize{Areas with positive and negative values of the Ricci scalar $\mathcal{R}_{\psi\theta}|_{\theta=0}$ of 2-dimensional $(\psi,\theta)$ section on the axis $\theta=0$ , for varying Monopole-Dipole distortion. In the left, we have $b_0=-3b_1$, $a_0=a_1=b_2=a_2=0$. In the right, we have $a_0=3a_1$, $b_0=b_1=b_2=a_2=0$. The white regions correspond to positive curvature, while the gray regions corresponds to negative curvature.}}
	\label{RicciPSiT0C1C2V}
\end{figure}

\begin{figure}[htp]
	\setlength{\tabcolsep}{ 0 pt }{\scriptsize\tt
		\begin{tabular}{ cccc }
			\includegraphics[width=15 cm]{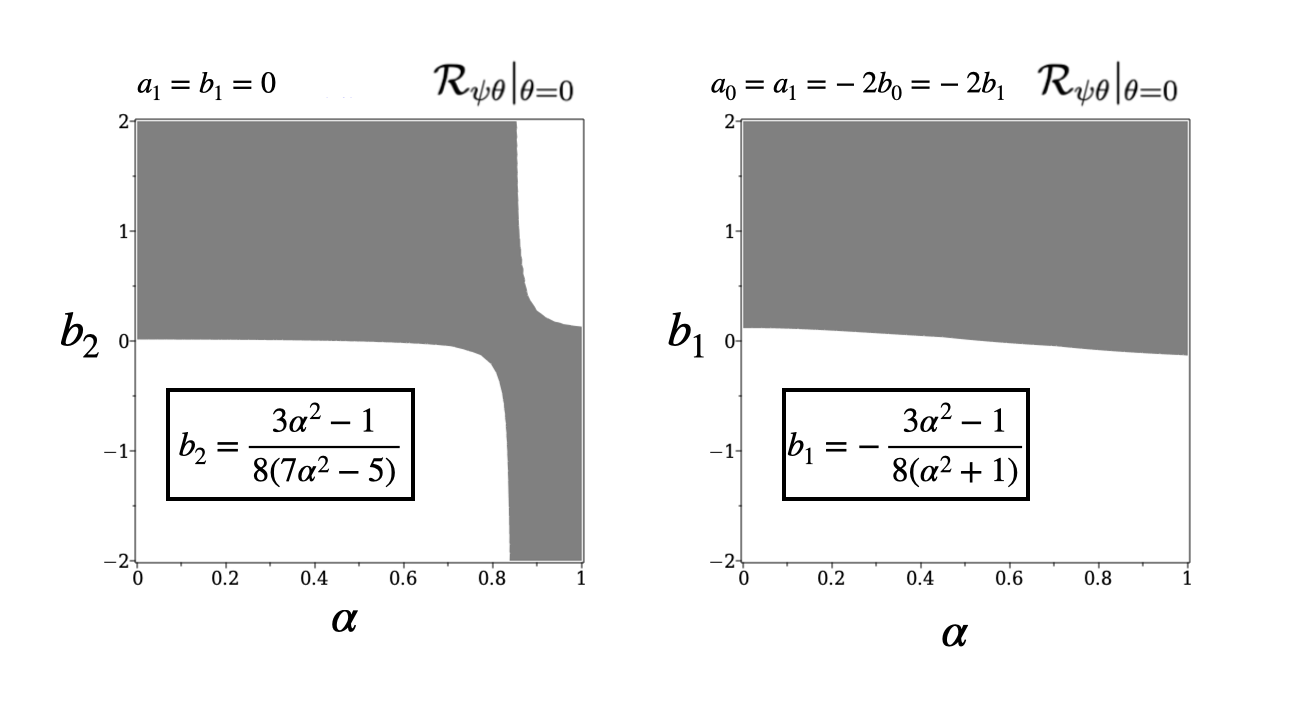} \\
	\end{tabular}}
	\caption{\footnotesize{Areas with positive and negative values of the Ricci scalar $\mathcal{R}_{\psi\theta}|_{\theta=0}$ of 2-dimensional $(\psi,\theta)$ section on the axis $\theta=0$, for varying distortion parameters. In the left, we have $a_1=b_1=0$ and $a_0=b_0=b_2=a_2$. In the right, we have $a_0=a_1=-2b_0=-2b_2$, and $b_2=a_2=0$. The white regions correspond to positive curvature, while the gray regions corresponds to negative curvature.}}
	\label{RicciPSIT0C3C42V}
\end{figure}

\begin{figure}[htp]
	\setlength{\tabcolsep}{ 0 pt }{\scriptsize\tt
		\begin{tabular}{ cccc }
			\includegraphics[width=15 cm]{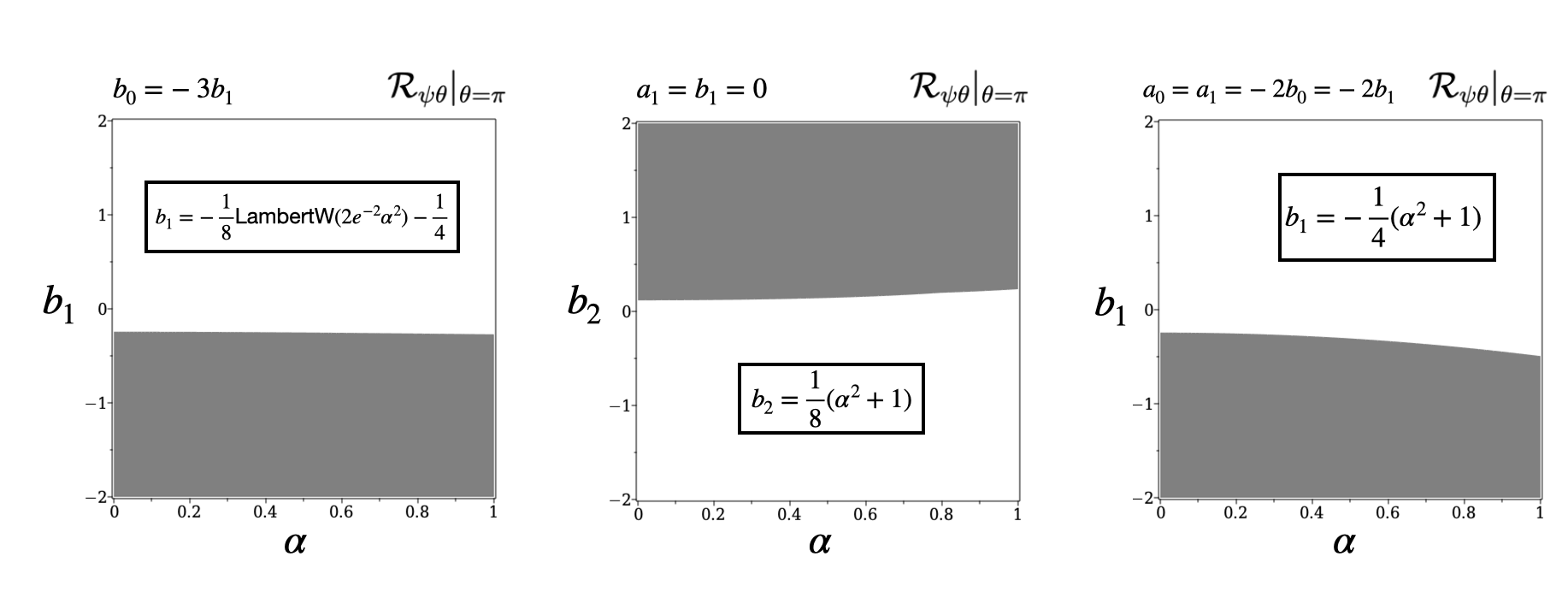} \\
	\end{tabular}}
	\caption{\footnotesize{Areas with positive and negative values of the Ricci scalar $\mathcal{R}_{\psi\theta}|_{\theta=\pi}$ of 2-dimensional $(\psi,\theta)$ section on the axis $\theta=\pi$ , for varying distortion parameter. In the left, we have $b_0=-3b_1$, $a_0=a_1=b_2=a_2=0$. In the middle, we have $a_1=b_1=0$ and $a_0=b_0=b_2=a_2$. On the right, we have $a_0=a_1=-2b_0=-2b_1$, and $b_2=a_2=0$. The white regions correspond to positive curvature, while the gray regions corresponds to negative curvature. In each case, we have also written the equation of the separating curve between the gray and white area.}}
	\label{RicciPSiTP.V}
\end{figure}

\begin{figure}[htp]
	\setlength{\tabcolsep}{ 0 pt }{\scriptsize\tt
		\begin{tabular}{ cccc }
			\includegraphics[width=15 cm]{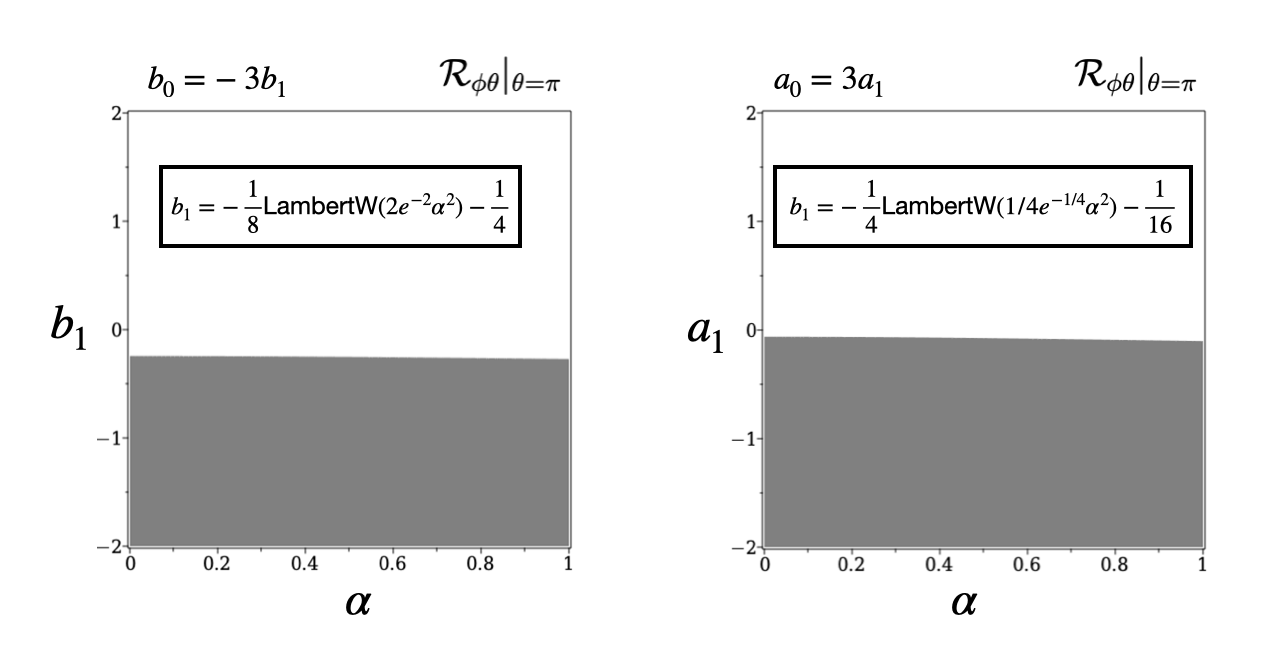} \\
	\end{tabular}}
	\caption{\footnotesize{Areas with positive and negative values of the Ricci scalar $\mathcal{R}_{\phi\theta}|_{\theta=\pi}$ of 2-dimensional $(\phi,\theta)$ section on the axis $\theta=\pi$ , for varying distortion parameter. In the left, we have $b_0=-3b_1$, $a_0=a_1=b_2=a_2=0$. In the right, we have $b_0=b_1=-2a_0=-2a_1$, and $b_2=a_2=0$. The white regions correspond to positive curvature, while the gray regions corresponds to negative curvature. In each case, we have also written the equation of the separating curve between the gray and white area.}}
	\label{RicciPHIPV}
\end{figure}

\begin{figure}[htp]
	\setlength{\tabcolsep}{ 0 pt }{\scriptsize\tt
		\begin{tabular}{ cccc }
			\includegraphics[width=15 cm]{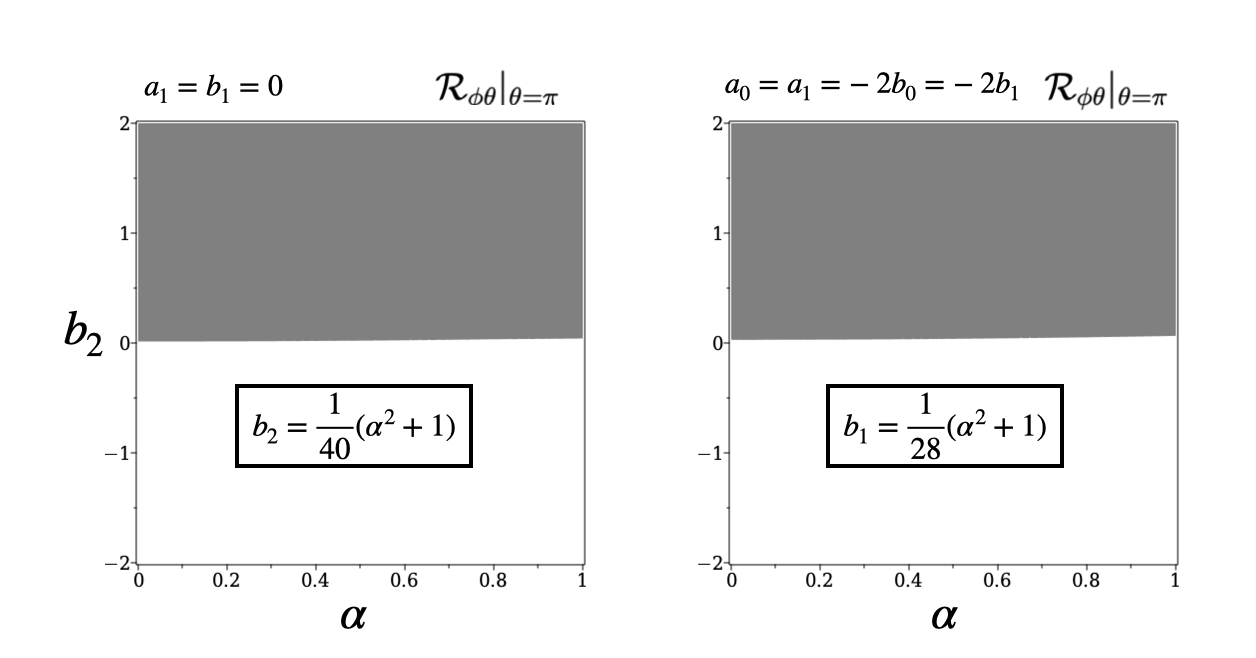} \\
	\end{tabular}}
	\caption{\footnotesize{Areas with positive and negative values of the Ricci scalar $\mathcal{R}_{\phi\theta}|_{\theta=\pi}$ of 2-dimensional $(\phi,\theta)$ section on the axis $\theta=\pi$ , for varying distortion parameter. In the left, we have $a_1=b_1=0$ and $a_0=b_0=b_2=a_2$. In the right, we have $a_0=a_1=-2b_0=-2b_1$, and $b_2=a_2=0$. The white regions correspond to positive curvature, while the gray regions correspond to negative curvature. In each case, we have also written the equation of the separating curve between the gray and white area.}}
	\label{RicciPHI2PV}
\end{figure}

In figures \ref{RicciPSiT0C1C2V}-\ref{RicciPHI2PV}, we have illustrated the regions where the Ricci scalar on the axes $\theta=0$ and $\theta=\pi$ is negative, as a function of the multiple moments.

For the Ricci scalar $\mathcal{R}_{\psi\theta}|_{\theta=\pi}$ of 2-dimensional $(\psi,\theta)$ section on the axis $\theta=\pi$, and distortion type II, i.e. $a_0=3a_1$, there exists no negative region. 

If we consider the Ricci scalar $\mathcal{R}_{\phi\theta}|_{\theta=0}$ of 2-dimensional $(\phi,\theta)$ section on the axis $\theta=0$ for the different cases of the distortion we find that for the monopole-dipole case  $b_0=-3b_1$, $a_0=a_1=b_2=a_2=0$, the Ricci scalar is always positive. This is also obvious from the simple expression of $\mathcal{R}_{\phi\theta}|_{\theta=0}$. For the case II, corresponding to $a_0=3a_1$, $b_0=b_1=b_2=a_2=0$, $\mathcal{R}_{\phi\theta}|_{\theta=0}$ is negative when $a_1>1/4$. In the quadrupole Case III, corresponding to $a_1=b_1=0$ and $a_0=b_0=b_2=a_2$, $\mathcal{R}_{\phi\theta}|_{\theta=0}$ is negative for $a_2>1/8$. For the case, $b_0=b_1=-2a_0=-2a_1$, and $b_2=a_2=0$, the Ricci scalar $\mathcal{R}_{\phi\theta}|_{\theta=0}$ is negative for $b_1<-1/8$. Figures \ref{RicciPSiT0C1C2V}-\ref{RicciPHI2PV} are consistent with figures \ref{Contour1}-\ref{Contour1_posV2} in illustrating for what ranges of parameters the isometric embedding of the sectional horizon surface into Euclidean space is possible. 

\subsection{Curvature of the horizon surface}
Consider the metric of the horizon in (\ref{HorizonM}) form. For a metric of a form
\be
ds^{2}=f(\theta)d\theta^{2}+g(\theta)d{\varphi}^{2}+h(\theta)d{\psi}^{2},\label{MetricS}
\ee
introduce the Gauss curvature tensor as follows
\be
\mathcal{K}_{ab}=-\mathcal{R}_{\hat{\alpha}\hat{\beta}\hat{\gamma}\hat{\delta}}e^{\hat{\alpha}}_{a}
e_{b}^{\hat{\beta}}e_{a}^{\hat{\gamma}}e_{b}^{\hat{\delta}}, 
\ee
where $e^{\hat{\alpha}}_{a}$, are 3 orthonormal vectors. For the metric ($\ref{MetricS}$) the directions of the coordinate lines $\psi$, $\theta$, and $\phi$ are eigen-vectors of $\mathcal{K}_{ab}$ and the corresponding
eigen-values are:
\ba
&&K_{\psi}=\frac{\mathcal{R}_{\theta\varphi\theta\varphi}}{fg},~~~
K_{\varphi}=\frac{\mathcal{R}_{\theta\psi\theta\psi}}{fh},~~~
K_{\theta}=\frac{\mathcal{R}_{\varphi\psi\varphi\psi}}{gh}.\label{RelationK}
\ea
We use 
\ba
&&\mathcal{R}_{\theta\varphi\theta\varphi}=\frac{g'(fg)'}{4fg}-\frac{1}{2}g'', \\
&&\mathcal{R}_{\theta\psi\theta\psi}=\frac{h'(fh)'}{4fh}-\frac{1}{2}h'', \\
&&\mathcal{R}_{\varphi\psi\varphi\psi}=-\frac{g'h'}{4f}, \label{RelationRS}
\ea
for the metric ($\ref{MetricS}$). For the 3-dimensional horizon surface components of its Ricci tensor are related to the Gauss curvatures of the sections as follows:
\ba
\mathcal{R}^{\varphi}_{~\varphi}=K_{\psi}+K_{\theta},~~
\mathcal{R}^{\psi}_{~\psi}=K_{\varphi}+K_{\theta},~~~
\mathcal{R}^{\theta}_{~\theta}=K_{\psi}+K_{\varphi},
\ea
where $K_{\psi}$, $K_{\phi}$, and $K_{\theta}$ are the curvatures of the 2D sections orthogonal to $\psi$, $\phi$ and $\theta$ lines, respectively. The Ricci scalar $\mathcal{R}$ and the trace of the square of the
Ricci tensor, $\mathcal{R}_{AB}\mathcal{R}^{AB}$, ($A$ and $B$ are $\psi$, $\theta$ and $\phi$) of the horizon surface are natural invariant measures of its intrinsic curvature, which are given as follows:
\ba
&&\mathcal{R}=\mathcal{R}^{\varphi}_{~\varphi}+\mathcal{R}^{\psi}_{~\psi}+\mathcal{R}^{\theta}_{~\theta}= 2(K_{\psi}+K_{\theta}+K_{\varphi})\nonumber\\
&&\mathcal{R}_{AB}\mathcal{R}^{AB}=(\mathcal{R}^{~\phi}_{\phi})^{2}+(\mathcal{R}^{~\psi}_{\psi})^{2}+(\mathcal{R}^{~\theta}_{\theta})^{2}.
\ea 
For 5-dimensional
vacuum static space-time, the Kretchman scalar of the spacetime calculated on the horizon is related to the square of Ricci tensor of the 3-dimendioal surface of the horizon surface by the following relation:
\ba 
\mathcal{K}=\mathcal{R}_{AB}\mathcal{R}^{AB}. 
\ea

\begin{figure}[!hbp]
	\begin{center}
		\includegraphics[width=16cm]{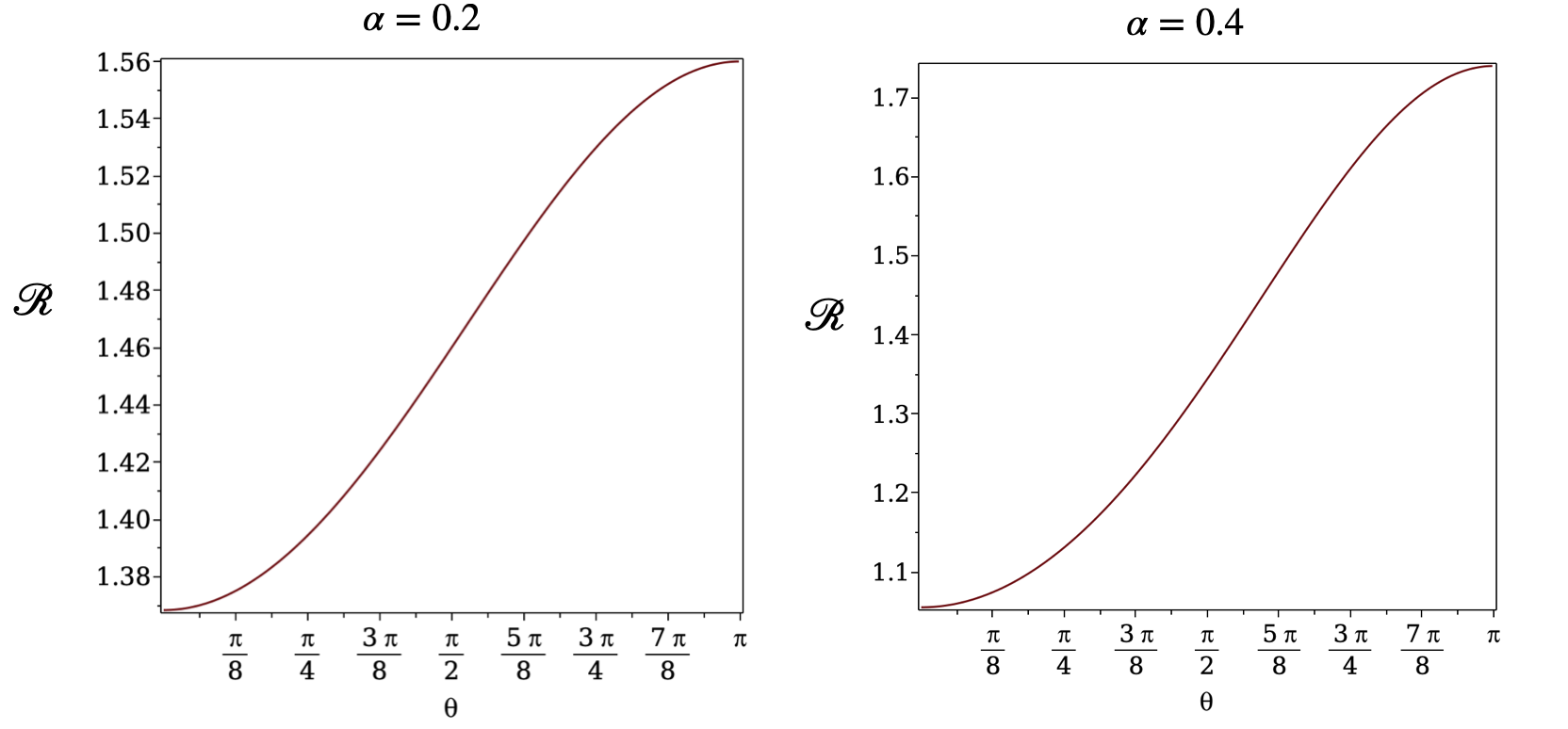}~\\
		\includegraphics[width=16cm]{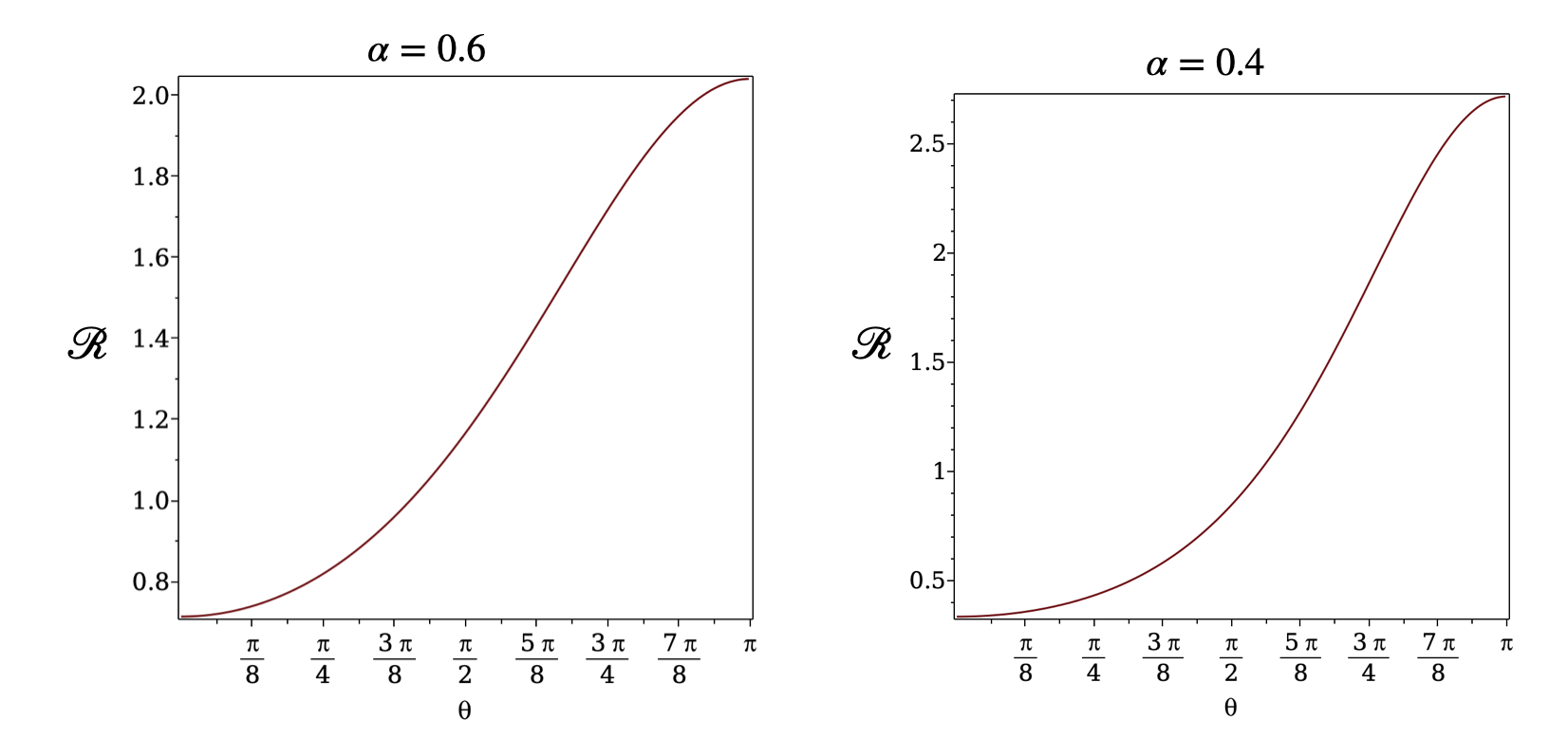}
		\caption{The Ricci scalar $\mathcal{R}$ of the horizon surface for the undistorted case. \label{Riccib0}}
	\end{center}
\end{figure}

The Ricci scalar of the horizon surface for the undistorted black Myers-Perry black hole is given by
\be
\mathcal{R}=-\frac{(\alpha^2+1)(\alpha^2 \cos^2(\frac{\theta}{2})-3)}{2(\alpha^2 \cos^2(\frac{\theta}{2})+1)^3}\, .
\ee
Figure \ref{Riccib0}, shows the Ricci scalar for the undistorted black hole for various values of $\alpha$. It can be seen that generally for all values of $\alpha$, curvature grows steadily from one pole $(\theta=0)$ to the opposite pole $(\theta=\pi)$, and increasing $\alpha$ significantly increases the Ricci scalar on the horizon. Furthermore, the black hole’s rotation introduces asymmetry into the horizon’s geometry, deforming it from a symmetric spherical shape into an oblate shape. As spin increases (from $\alpha=0.2$ up to $\alpha=0.9$), the difference between the minimum and maximum curvature significantly grows.

\begin{figure}[!hbp]
	\begin{center}
		\includegraphics[width=8cm]{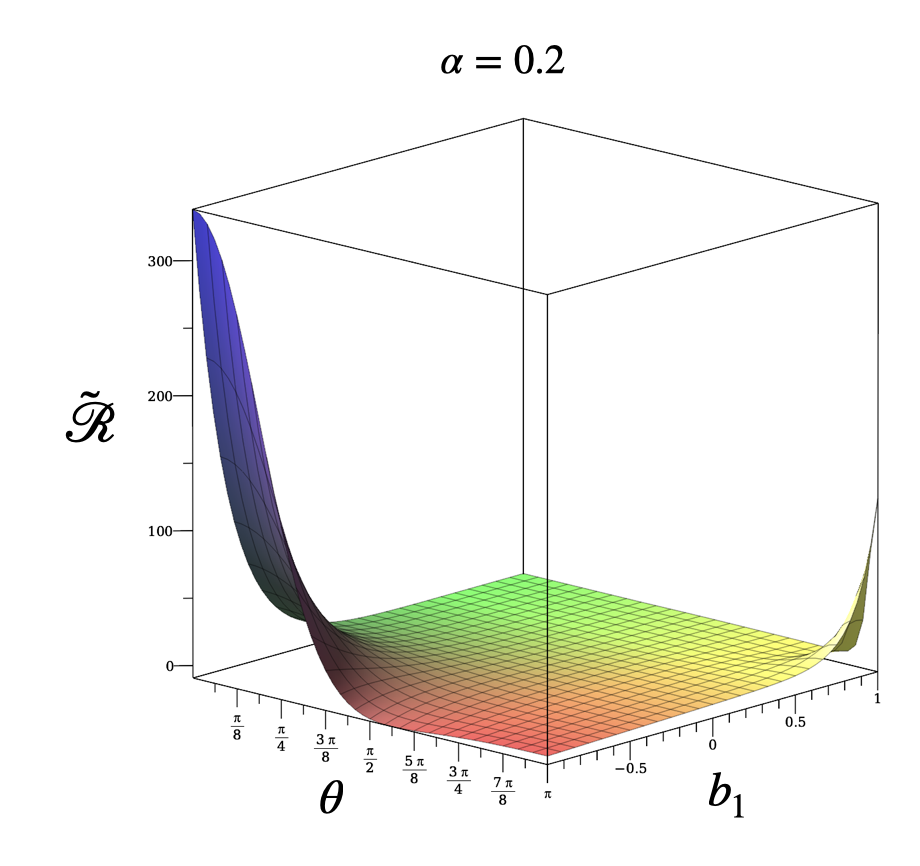}~
		\includegraphics[width=8cm]{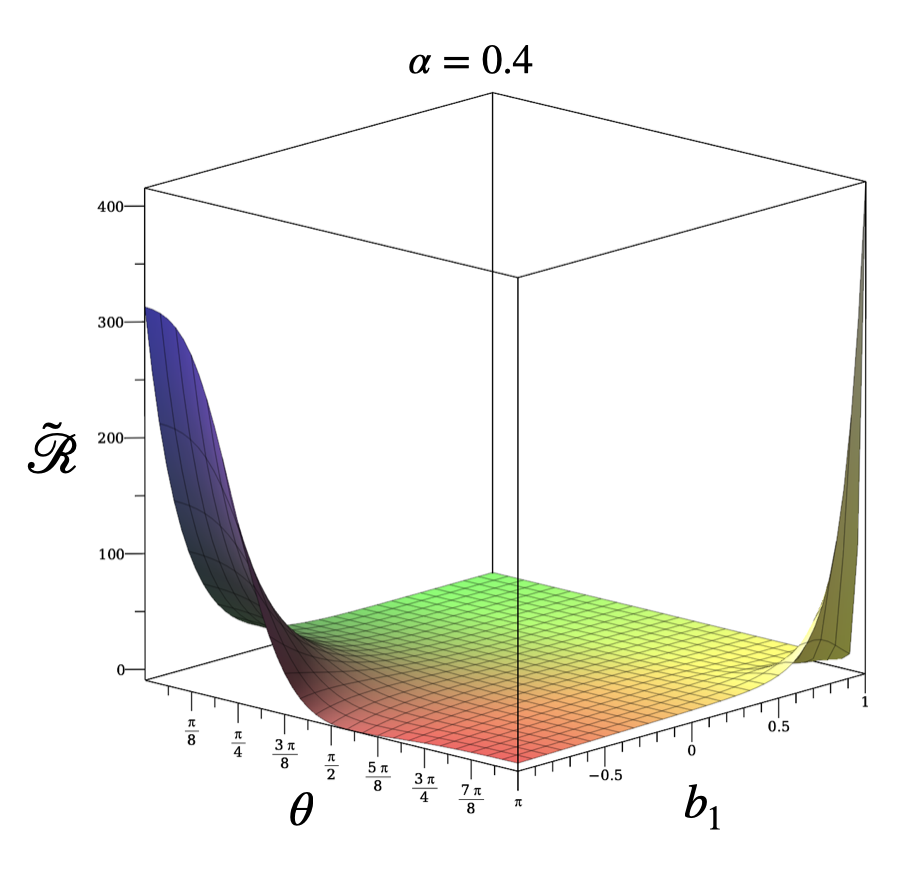}~\\
		\includegraphics[width=8cm]{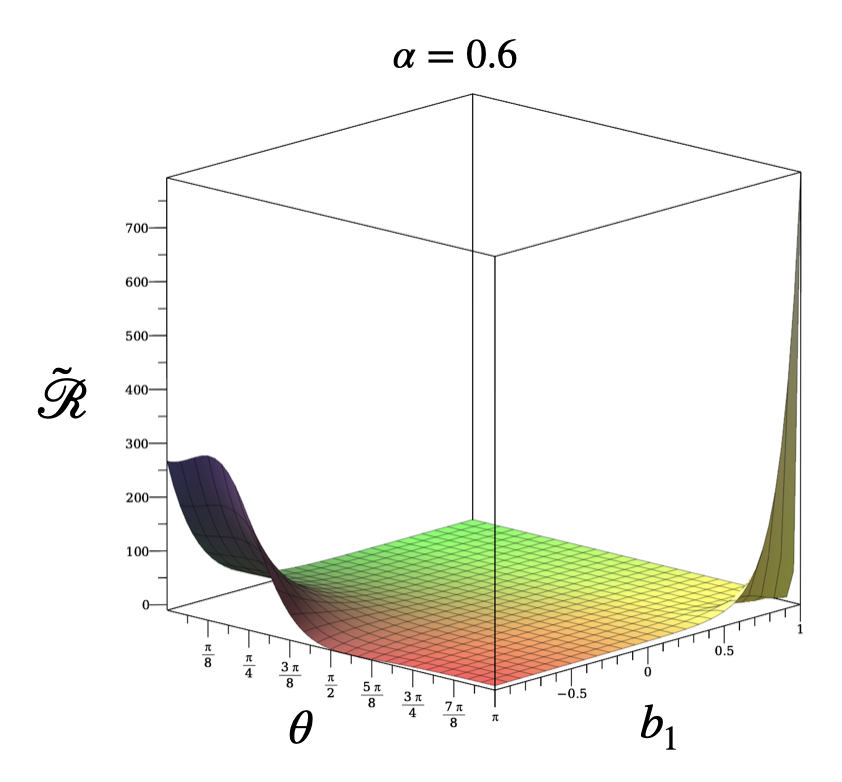}~
		\includegraphics[width=8cm]{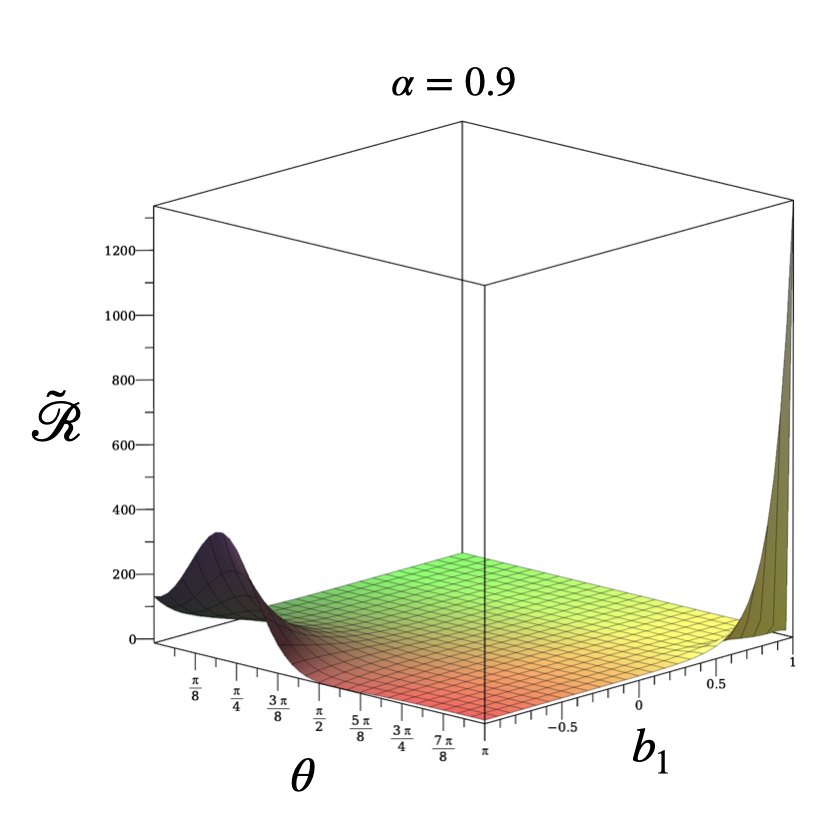}
		\caption{The normalized Ricci scalar $\tilde{\mathcal{R}}$ of the horizon surface for $b_0=-3b_1$, and $b_2=a_0=a_1=a_2=0$ \label{Riccib1}}
	\end{center}
\end{figure}

\begin{figure}[!hbp]
	\begin{center}
		\includegraphics[width=8cm]{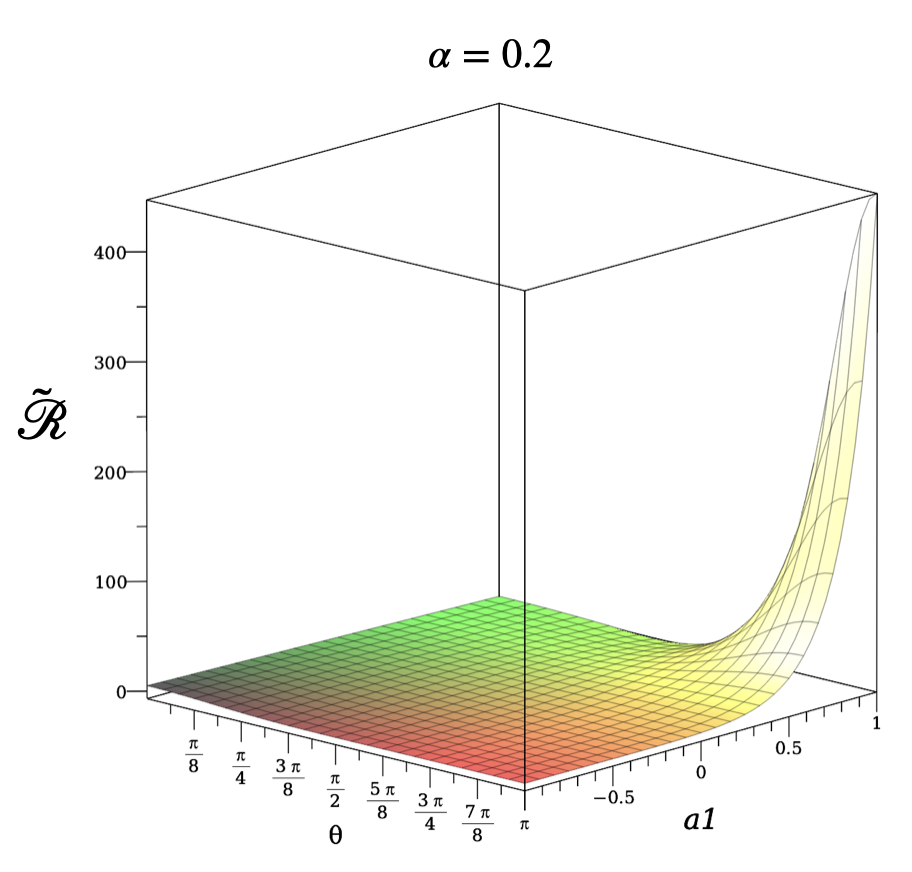}~
		\includegraphics[width=8cm]{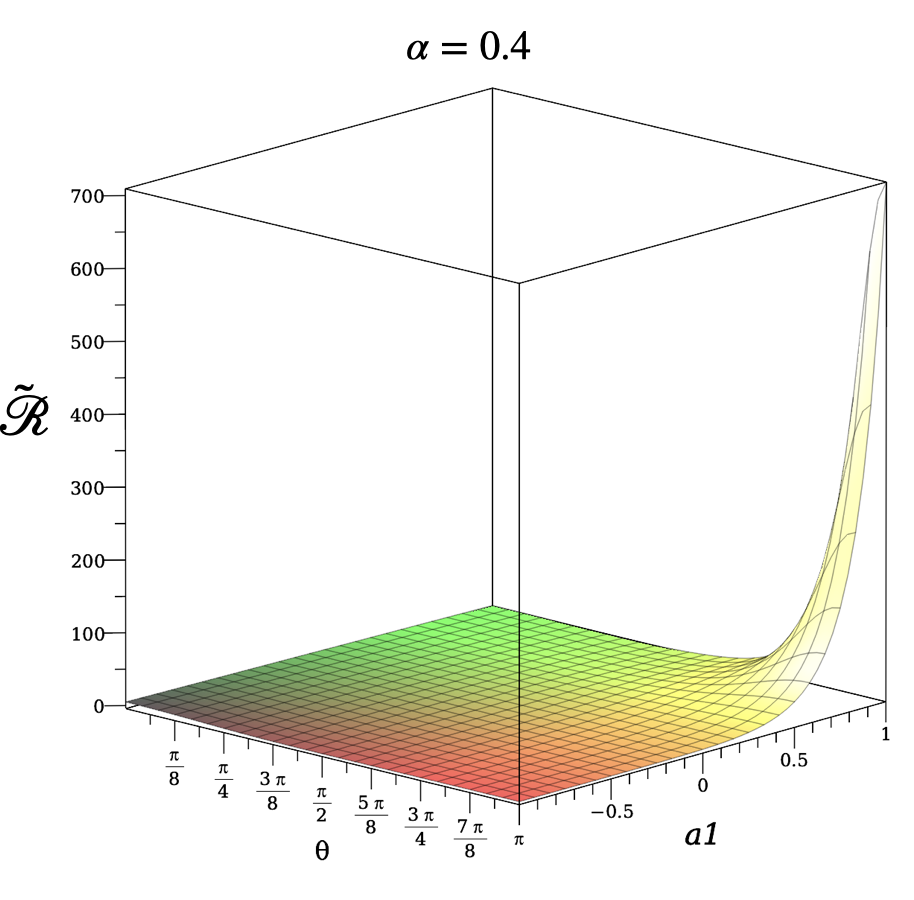}~\\
		\includegraphics[width=8cm]{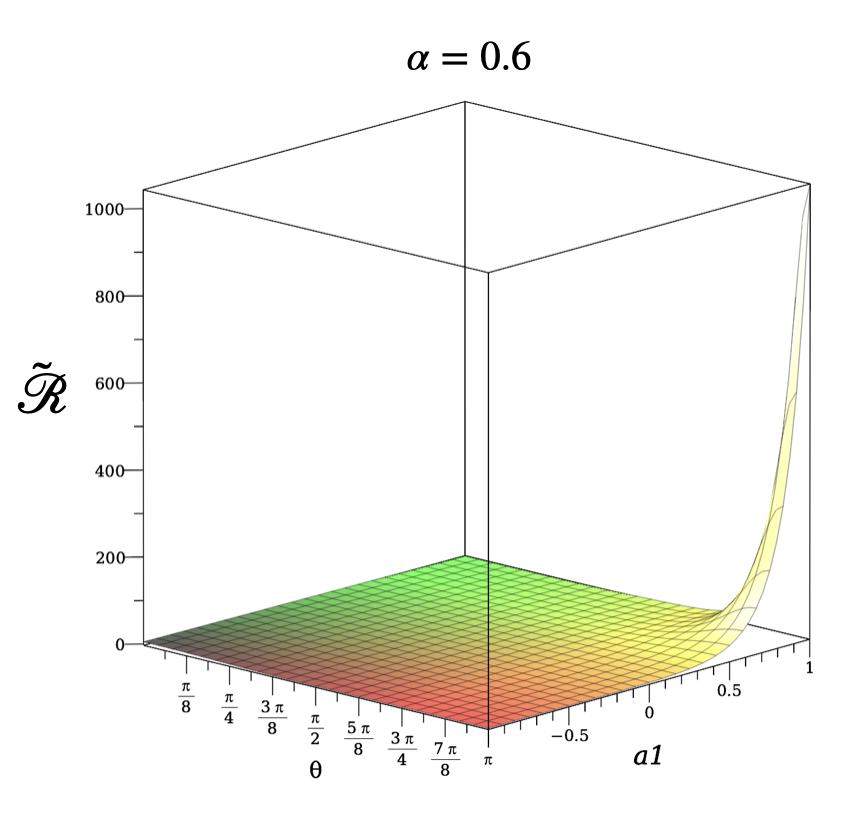}~
		\includegraphics[width=8cm]{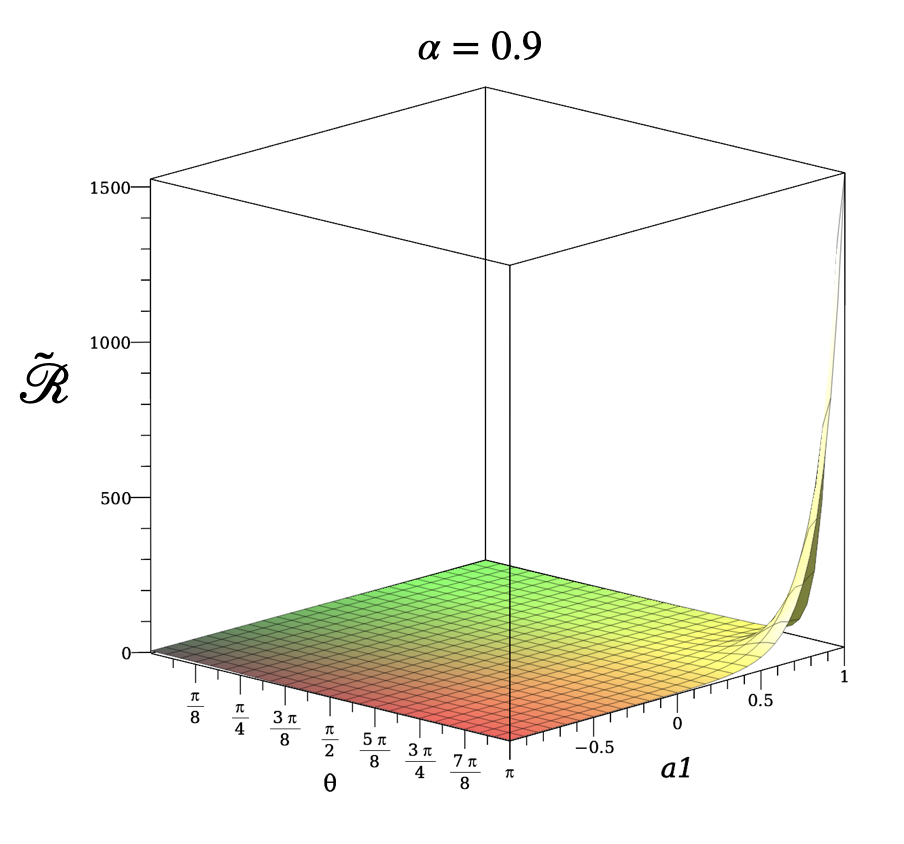}
		\caption{The normalized Ricci scalar $\tilde{\mathcal{R}}$ of the horizon surface for $a_0=3a_1$, and $b_0=b_1=b_2=a_2=0$ \label{Riccib2}}
	\end{center}
\end{figure}

\begin{figure}[!hbp]
	\begin{center}
		\includegraphics[width=8cm]{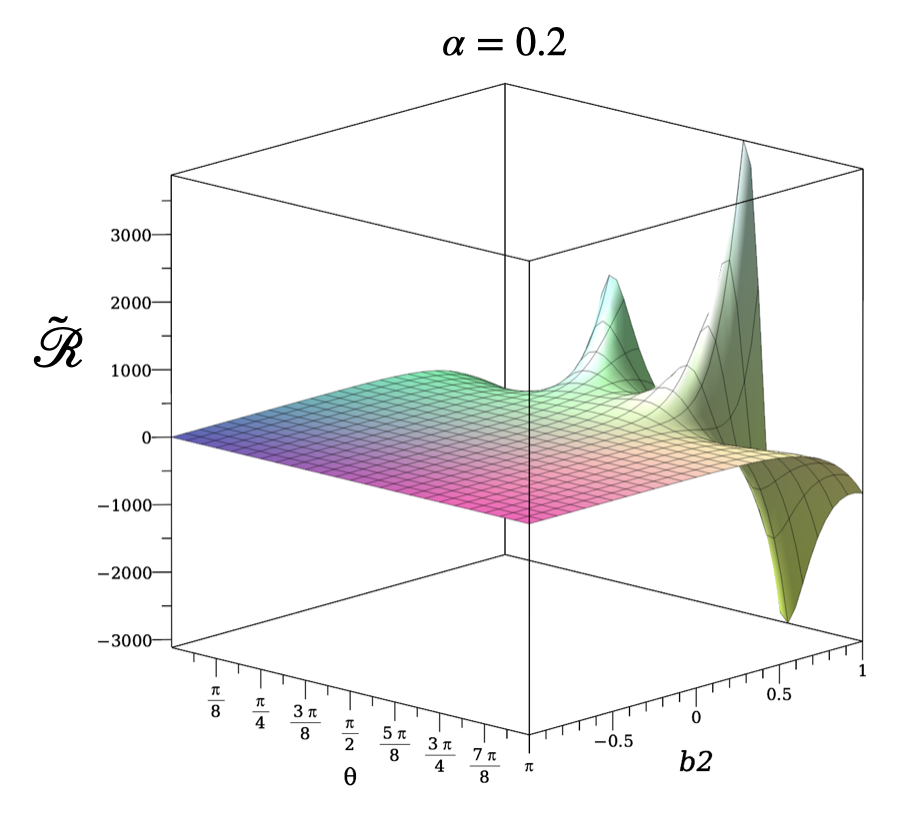}~
		\includegraphics[width=8cm]{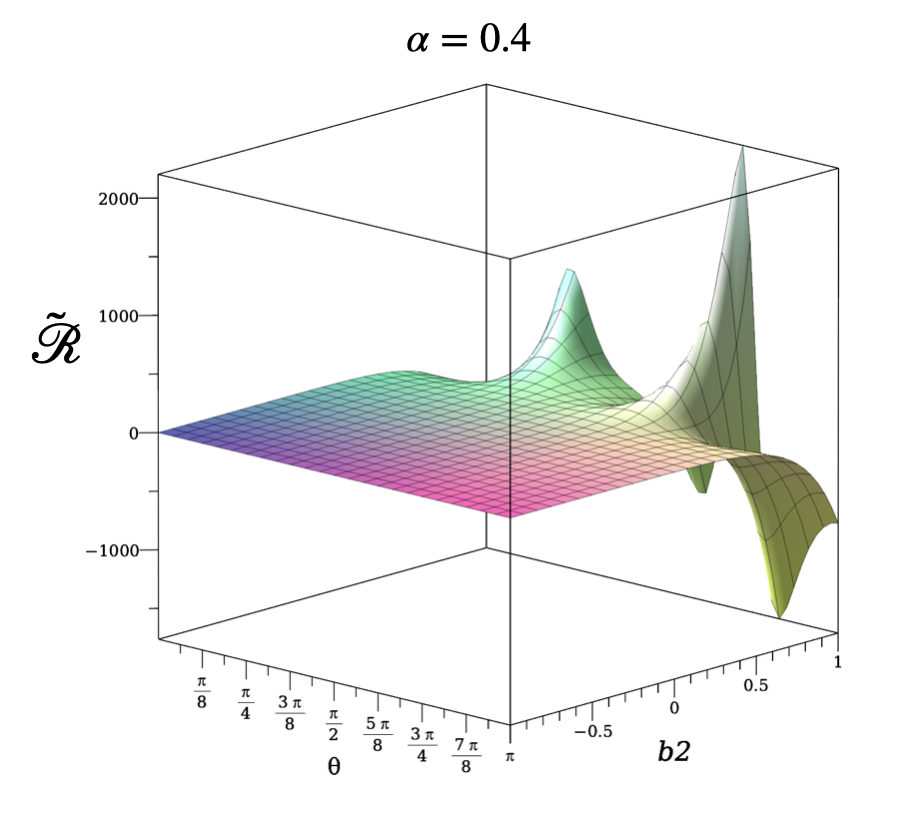}~\\
		\includegraphics[width=8cm]{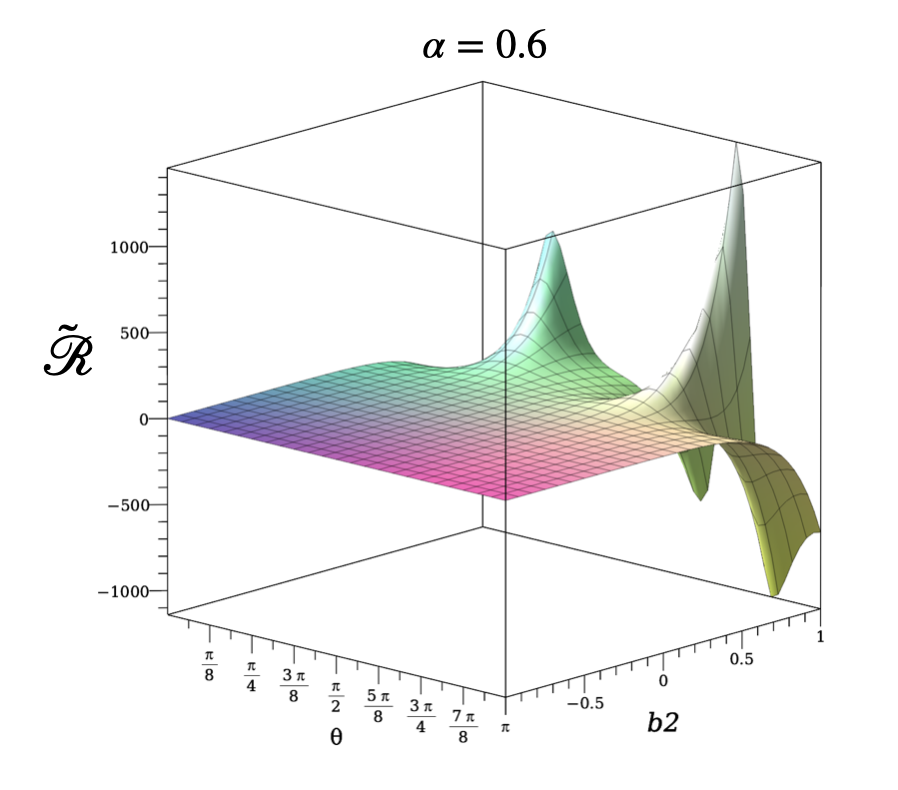}~
		\includegraphics[width=8cm]{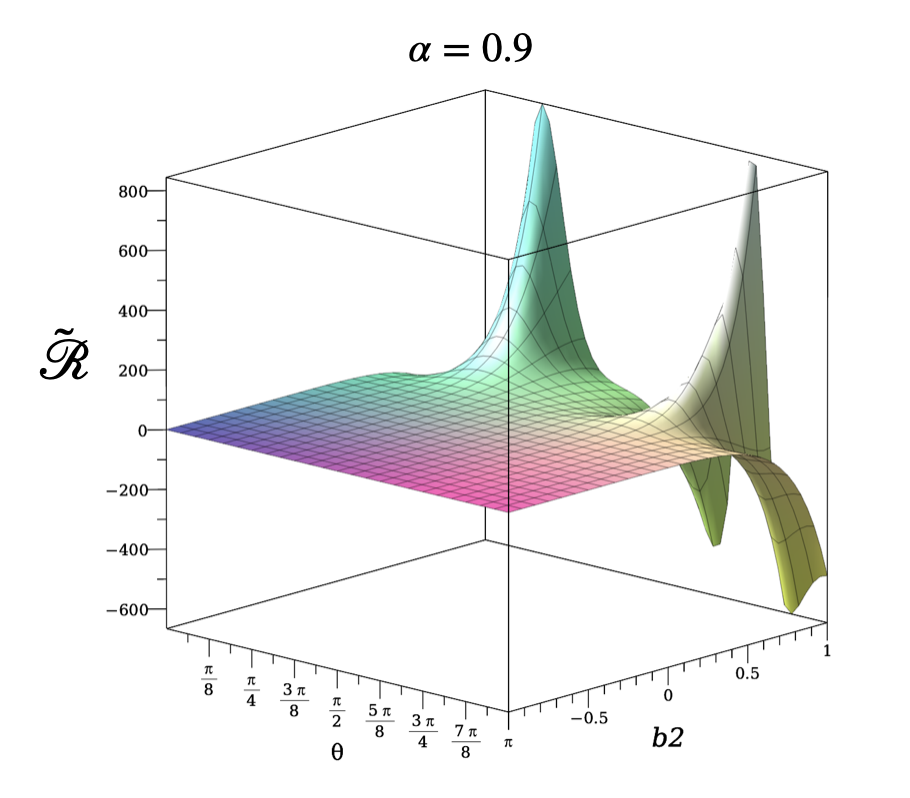}
		\caption{The normalized Ricci scalar $\tilde{\mathcal{R}}$ of the horizon surface for $a_1=b_1=0$, and $a_0=b_0=b_2=a_2$ \label{Riccib3}}
	\end{center}
\end{figure}

\begin{figure}[!hbp]
	\begin{center}
		\includegraphics[width=16cm]{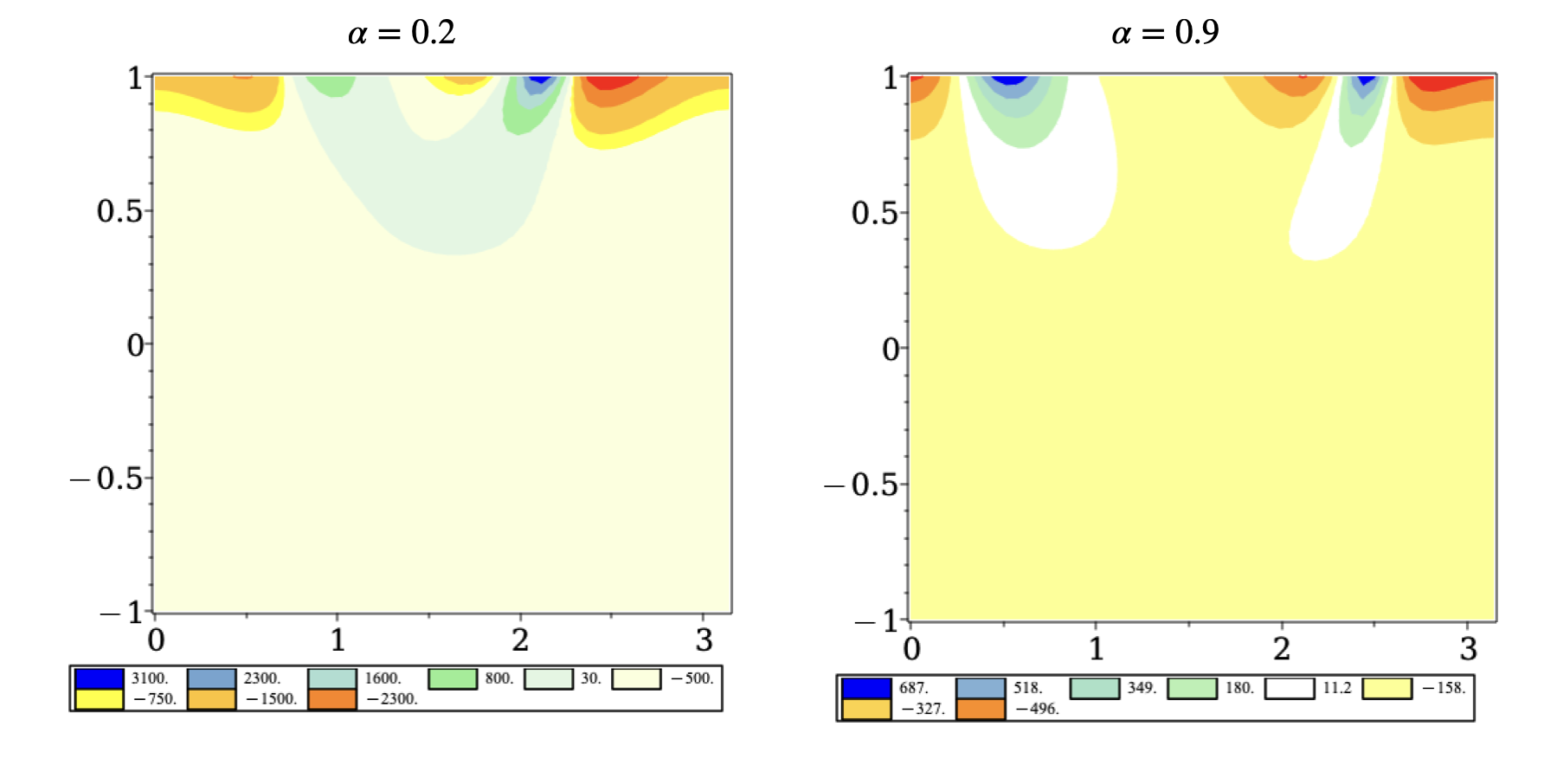}~
		\caption{Contour plots for the normalized Ricci scalar $\tilde{\mathcal{R}}$ of the horizon surface for $a_1=b_1=0$, and $a_0=b_0=b_2=a_2$. Color labels show regions with values larger than shown in the legend.\label{Riccib3B}}
	\end{center}
\end{figure}

\begin{figure}[!hbp]
	\begin{center}
		\includegraphics[width=16cm]{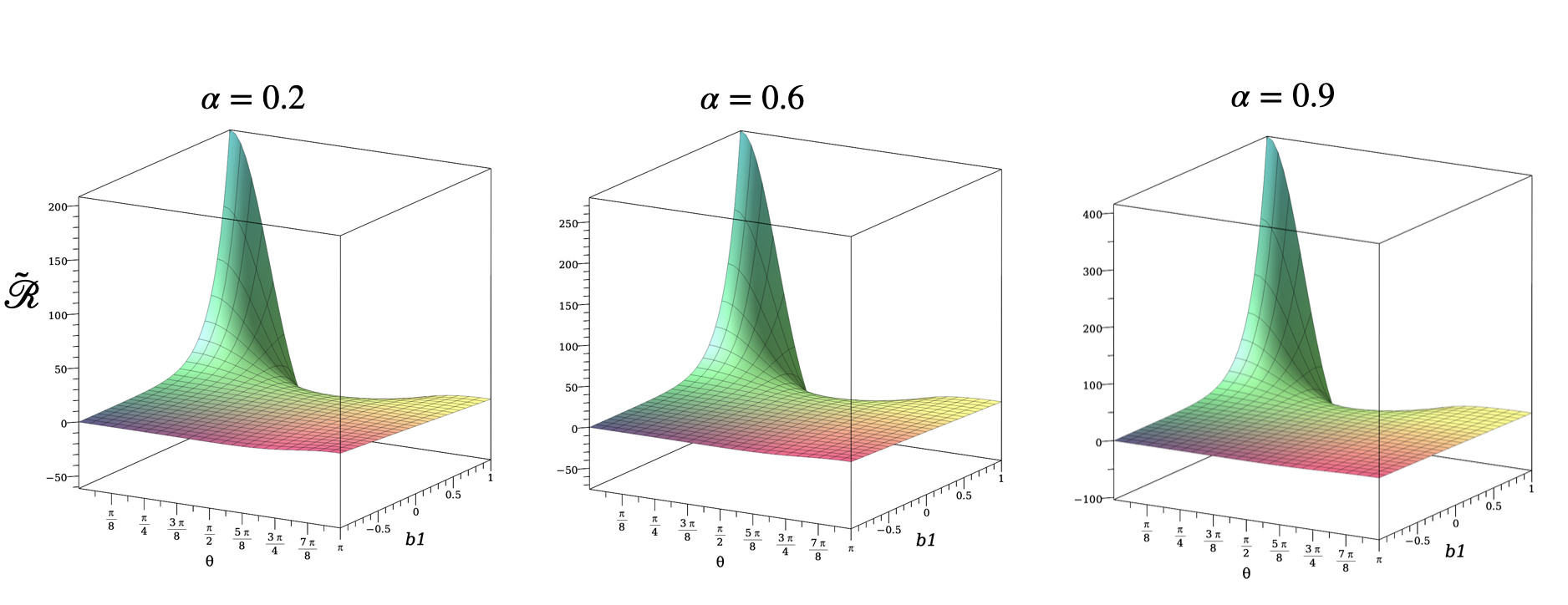}~
		\caption{The normalized Ricci scalar $\tilde{\mathcal{R}}$ of the horizon surface for $a_2=b_2=0$, and $a_0=a_1=-2b_0=-2b_1$ \label{Riccib4}}
	\end{center}
\end{figure}

\begin{figure}[!hbp]
	\begin{center}
		\includegraphics[width=16cm]{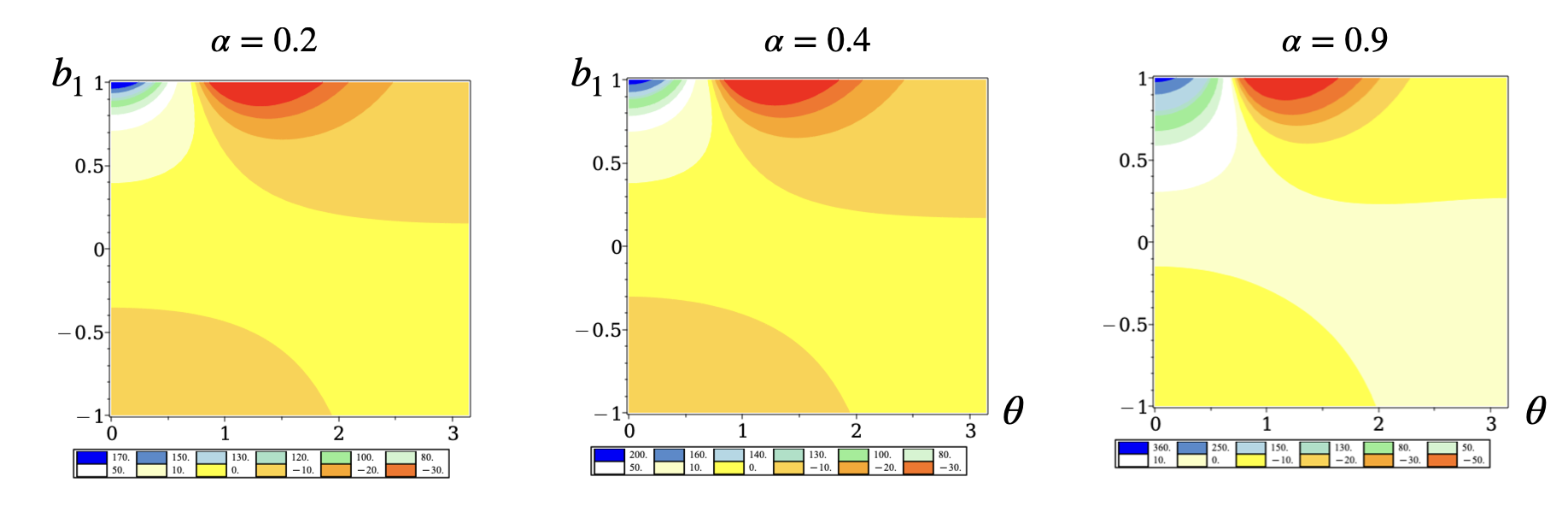}~
		\caption{Contour plots for the normalized Ricci scalar $\tilde{\mathcal{R}}$ of the horizon surface for $a_2=b_2=0$, and $a_0=a_1=-2b_0=-2b_1$. Color labels show regions with values larger than shown in the legend. \label{Riccib5}}
	\end{center}
\end{figure}

In figures \ref{Riccib1}-\ref{Riccib5}, we have considered the normalized Ricci scalar of the horizon surface $\tilde{\mathcal{R}}$ defined as the ratio of the Ricci scalar of the distorted black hole to the Ricci scalar of the undistorted black hole. Note that in the Ricci scalar of the horizon surface for the distorted black hole, $\mathcal{R}$, monopoles $a_0$ and $b_0$ appear as a scaling factor. To find the Ricci scalar that is independent of the monopole, divide $\mathcal{R}$ by $\exp(a_0+b_0)$.

Figure \ref{Riccib1} presents the normalized Ricci scalar for four different rotation parameters, $\alpha = 0.2, 0.4, 0.6,$ and $0.9$, shown in panels (a)–(d), respectively. In all plots, we have chosen $ b_{0}=-3 b_{1}$ and set the remaining distortion parameters $b_{2}=a_{0}=a_{1}=a_{2}=0$, so that only a pure dipole is present. Here $\tilde{\mathcal{R}}$ remains almost flat (i.e.\ very close to unity) for moderate distortions $b_{1}\le 0.5$. . As $\alpha$ increases, the overall height of the curvature bump grows. Moreover, a secondary ridge begins to form for negative $b_{1}$ in the southern hemisphere $(\theta<0)$, showing that moderate spin allows the horizon to develop more structure in response to a dipole source. “For large $\alpha$, the horizon responds strongly to any external dipole source. Rather than remaining smooth, it develops localized curvature lobes—regions where the Ricci scalar rises well above its average value.
In figure \ref{Riccib2}, positive $a_1$ distortions produce strong curvature peaks at one hemisphere (around $\theta=\pi$), intensifying with increasing $\alpha$. In figure \ref{Riccib3}, a quadrupole distortion is represented, which produced a two-lobe pattern on the horizon curvature, symmetric about the equator. Their heights grow with both the distortion parameter $b_2$ and $\alpha$. It introduces more complicated curvature behavior, creating multiple peaks and valleys. Curvature changes sign across the parameter range, indicating the horizon's regions of positive and negative curvature.
In figure \ref{Riccib4}, with combined parameters, the Ricci scalar exhibits a prominent, sharp peak at the central angular region, increasing with $\alpha$. It shows how specific combinations of distortion parameters can significantly alter the horizon geometry.

\begin{figure}[!hbp]
	\begin{center}
		\includegraphics[width=16cm]{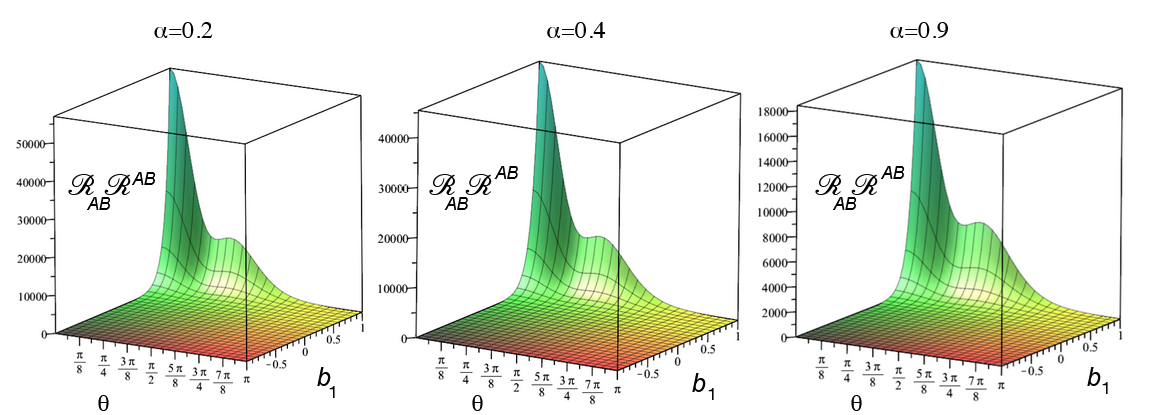}~
		\caption{The trace of the square of the Ricci
			tensor $\mathcal{R}_{AB}$ of the horizon surface for $a_2=b_2=0$, and $a_0=a_1=-2b_0=-2b_1$ \label{traceRicci}}
	\end{center}
\end{figure}

The case $a_1=-2b_1$ is of special interest. In this case, we have a bumpy deformed black hole horizon that feels the presence of the external sources, however, the ergosphere is unaffected by the presence of external sources.
For this reason, in Figs. \ref{Riccib4}-\ref{traceRicci}, we have further focused on this case. 
In Figure \ref{traceRicci}, we examined the trace of the square of the horizon’s Ricci tensor—as a function of $\theta$ and the dipole parameter $b_1$, for three values of $\alpha=0.2,0.4,0.9$. A very sharp, high amplitude peak appears when $b \to 1$ at the pole,$\theta\to0$. This is where the Ricci tensor squared invariant blows up, indicating an intense, highly localized curvature concentration. This effect becomes stronger and more localized as the $\alpha$ increases. 
These plots illustrate how external gravitational fields alter the horizon structure of rotating Myers-Perry black holes. Distorted horizons develop intense localized curvature and tidal features, amplified by higher values of $\alpha$. While undistorted black holes present smoother, symmetric curvature distributions, distortions break this symmetry.

\newpage
\section{Summary}
In this paper, we studied the effects of external distortions on a five-dimensional Myers-Perry black hole. 

The primary focus was on analyzing how distortions induced by external fields affect the horizon geometry of the five-dimensional Myers-Perry black hole.  The horizon surface is deformed due to the external fields such as monopole, dipole, and quadrupole distortions. The analysis reveals that distortions alter the shape, curvature, and embedding properties of the horizon surface, leading to significant deviations from the undistorted Myers-Perry geometry. 

Furthermore, an embedding analysis of the horizon's sections into three-dimensional Euclidean space has been performed.
This study shows that for an undistorted Myers-Perry black hole, the rotational carves $ (\psi,\theta) $ are more oblate for larger values of $ \alpha $, and there are values of $ \alpha $ which the embedding of the horizon surface into the Euclidean space is impossible for them. Furthermore, for  $ (\phi,\theta) $ section of the horizon surface, the larger values of $ \alpha $ lead to more prolate horizon surface. For the distorted case and monopole-dipole case the larger positive values of $b_1$ or $a_1$ correspond to a more prolate horizon surface.

In the presence of distortion fields, the possibility of this embedding depends on both the value of $ \alpha $ and the distortion parameters. 
This study shows distortions can render embedding in three-dimensional Euclidean space impossible for certain parameter ranges. Figures \ref{Contour1}-\ref{Contour1_posV2} show the ranges of distortion paramters and $\alpha$, where the embedding into Euclidean space is possible.

In the absence of external fields, the intrinsic scalar curvature $\mathcal{R}$ of a Myers–Perry horizon is a smooth, monotonic function of polar angle $\theta$, minimal at ($\theta=0$) and maximal at ($\theta=\pi$). Increasing the rotation parameter  $\alpha$ uniformly increases $\mathcal{R}$. In the presence of external fields, the intrinsic curvature of a rotating Myers–Perry horizon departs from its smooth, pole‑to‑pole gradient in the undistorted case. A dipole distortion with $b_0=-3b_1$ lead to two curvature lobes in opposite hemispheres, while a dipole distortion with $a_0=3a_1$ produces a single dominant peak.
The higher‑order (quadrupole) terms and mixed monopole–dipole combinations generate additional lobes or sharply localized peaks. Furthermore, the trace of the squared Ricci tensor, $\mathcal{R}_{AB}\mathcal{R}^{AB}. $ further reveals where tidal stresses concentrate most intensely, sharp peaks in $\mathcal{R}_{AB}\mathcal{R}^{AB} $ coincide with extrema of the Ricci scalar, and both quantities grow with the rotation parameter $\alpha.$ Increasing positive values of distortion parameter $b_1$ induces increasing variation of curvature invariants across the horizon and in general making the horizon much less smooth. In addition, curvature can increase orders of magnitude in the horizon due to the distortions.

\paragraph{Acknowledgements} M.T. acknowledges the support of the NSERC Discovery Grant 2018-
	04887 and NSERC Discovery Grant 2018-04873. S.A. acknowledges the hospitality of the California Institute of Technology’s Department of Theoretical Astrophysics, where part of this work was conducted.
	
	% --- References ---

\end{document}